\documentclass[12pt]{article}
\usepackage{graphicx,amsmath,amsfonts,amssymb
}
\usepackage{cite}
\usepackage[normalem]{ulem}
\newcommand*{\soutc}[1]{\unskip}
\newcommand*{\soutb}[1]{\unskip}
\newcommand{\souta}[1]{\unskip}

\usepackage[toc,page]{appendix}
\usepackage{color}
\usepackage{comment}
\usepackage[nottoc
]{tocbibind}
\DeclareMathOperator{\sech}{sech}

\def\be{\begin{equation}}
\def\ee{\end{equation}}

\usepackage{comment}
\usepackage{setspace}

\def\vec#1{\mbox{\boldmath $#1$}} 

\def\cyan{\textcolor{black}}
\def\blue{\textcolor{black}}

\def\be{\begin{equation}}
\def\ee{\end{equation}}

\def\be{\begin{equation}}
\def\ee{\end{equation}}

\def\be{\begin{equation}}
\def\ee{\end{equation}}

\def\red{\textcolor{black}}
\def\blue{\textcolor{black}}
\def\sky{\textcolor{black}}
\def\fish{\textcolor{black}}
\def\cat{\textcolor{black}}
\def\dog{\textcolor{black}}
\def\bl{\textcolor{black}}
\def\umi{\textcolor{black}}
\def\sea{\textcolor{black}}
\def\ao{\textcolor{black}}
\def\sora{\textcolor{black}}

\def\cy{\textcolor{black}}
\newcommand{\maru}[1]{\left( #1 \right)}
\newcommand{\kaku}[1]{\left[ #1 \right]}
\newcommand{\nami}[1]{\left\{ #1 \right\}}

\def\f{\frac}

\begin{document}

\begin{titlepage}
\begin{flushright}
RESCEU-28/16

\end{flushright}
\begin{center}


\vskip .5in

{\Large \bf
Supermassive black holes \red{formed by} direct collapse of inflationary perturbation\red{s}
}
\vskip .45in

{\large
Tomohiro Nakama$^{1}$,
Teruaki Suyama$^{2}$
and Jun'ichi Yokoyama$^{2,3,4}$
}

\vskip .45in%
{\em
$^1$
   Department of Physics and Astronomy, \\ Johns Hopkins University, Baltimore, Maryland 21218,
USA}\\
{\em
$^2$
   Research Center for the Early Universe (RESCEU), Graduate School
  of Science,\\ The University of Tokyo, Tokyo 113-0033, Japan
    }\\
    {\em
$^3$
   Department of Physics, Graduate School of Science,\\ The University of Tokyo, Tokyo 113-0033, Japan
    }\\
    {\em
$^4$
   Kavli Institute for the Physics and Mathematics of the Universe (Kavli IPMU),
\\ The University of Tokyo, Kashiwa, Chiba 277-8568, Japan
    }

\end{center}

\vskip .4in

\begin{abstract}
We propose a mechanism of producing a new type of primordial perturbations
that collapse to primordial black holes whose mass can be as large as necessary for them to grow to the 
supermassive black holes observed at high redshifts, without contradicting COBE/FIRAS upper limits on cosmic microwave background (CMB) spectral distortions.
In our model, the observable Universe consists of two kinds of many small patches which
experienced different expansion histories during inflation.
Primordial perturbations large enough to form primordial
black holes are realized on patches that experienced more Hubble expansion than the others.
By making these patches the minor component, 
the rarity of supermassive black holes can be explained.  
On the other hand, most regions of the Universe experienced the standard history and, hence, only have standard almost-scale-invariant adiabatic perturbations confirmed by observations of CMB or large-scale structures of the Universe. 
Thus, our mechanism can evade the constraint from the nondetection 
of the CMB distortion set by the COBE/FIRAS measurement.
Our model predicts the existence of supermassive black holes even at redshifts much higher than those observed. 
Hence, our model can be tested by future observations peeking into the higher-redshift Universe.
\end{abstract}
\end{titlepage}
\section{Introduction}
Observations have revealed the existence of supermassive black hole\red{s (}SMBHs) 
of about $10^9 M_\odot$ at high redshifts $z=6\sim 7$\sora{. S}o far, about 40 quasars, 
which are thought to be SMBHs \red{blazing} by accreting the surrounding gas, have been discovered
\cite{Fan:2003wd,Willott:2003xf,Kurk:2007qk,Jiang:2007qm,Willott:2007rm,Jiang:2007fz,Jiang:2009sd,Willott:2009wv,Willott:2010yu,DeRosa:2011bca,Mortlock:2011va,DeRosa:2013iia,Banados:2014lia}.
In particular, a quasar indicating a SMBH as massive as $1.2\times 10^{10}M_\odot$ was discovered recently \cite{Wu:2015bn}. 
Until now, there has been no established astrophysical explanation
of 
why such massive black holes (BHs) already existed at such high redshifts 
when the age of the Universe \sora{was} less than \red{a} billion years (see, e.g., \cite{Dokuchaev:2007mf,Volonteri:2010wz,Sesana:2011qi,Treister:2011yi,Haiman:2012ic}
for reviews of SMBHs in the high\cyan{-}redshift universe).

In light of this situation, it is intriguing to consider a possibility
that the observed SMBHs are primordial black holes (PBHs) that formed
in the very early Universe when the Universe was still dominated by radiation \cite{Zeldovich:1967ei}.
If some region has \umi{a curvature }perturbation of the order of unity,
this region undergoes gravitational collapse shortly after the size of the region 
becomes comparable to the Hubble horizon \cite{Hawking:1971ei,Carr:1974nx}. 
\red{Typically, the m}ass of the resultant black hole is \red{roughly} equal to the horizon mass at \sora{formation}. 
\bl{Since} the formation time of PBHs can be related to the \bl{comoving wave number $k$} of the perturbations collapsing to PBHs, 
\sora{their mass} can also be related \umi{to it} \bl{as $M_{\mathrm{PBH}}\sim 2\times10^{13}M_\odot (k/\mathrm{Mpc}^{-1})^{-2}$}. 
At first sight, \red{the} desired amount of PBHs of the desired mass, i.e., as large as necessary to grow to the order of $10^{9}M_\odot$ by $z\sim 6,7$, seems to be 
realized \soutb{by just preparing} \fish{just by a} moderate probability of primordial perturbation\red{s} of order unity 
at the corresponding (comoving) scale. 
Such perturbation\red{s} can indeed be realized in some inflation models \cite{PhysRevD.50.7173,GarciaBellido:1996qt,Yokoyama:1995ex,Kawasaki:1997ju,Yokoyama:1998pt,Kawasaki:1998vx,Yokoyama:1998qw,Taruya:1998cz,Kanazawa:2000ea,Kawaguchi:2007fz,Saito:2008em,Alabidi:2009bk,Frampton:2010sw,Suyama:2011pu,Kohri:2012yw,Kawasaki:2012kn,Kawasaki:2012wr,Suyama:2014vga},
\red{\bl{though} the sufficient f}ormation of such black holes does not happen in the standard cosmology in which
primordial perturbations are almost scale invariant and Gaussian \cite{Carr:1994ar}. 
\bl{The} approximate scale invariance and Gaussianity of the primordial
perturbation are observationally confirmed at large scales, namely, \red{the scales relevant to observations of the cosmic microwave background (CMB)} 
\red{(f}or recent Planck results, see \cite{Planck:2015xua,Ade:2015ava}) \cyan{or large-scale structures of the Universe}.
\umi{Yet} these properties could be largely violated on much shorter scales, 
\red{including the} scales corresponding to \red{the PBHs relevant to the seeds of SMBHs considered in this paper}.

\begin{figure}[t]
\begin{center}
\includegraphics[width=13cm,keepaspectratio,clip]{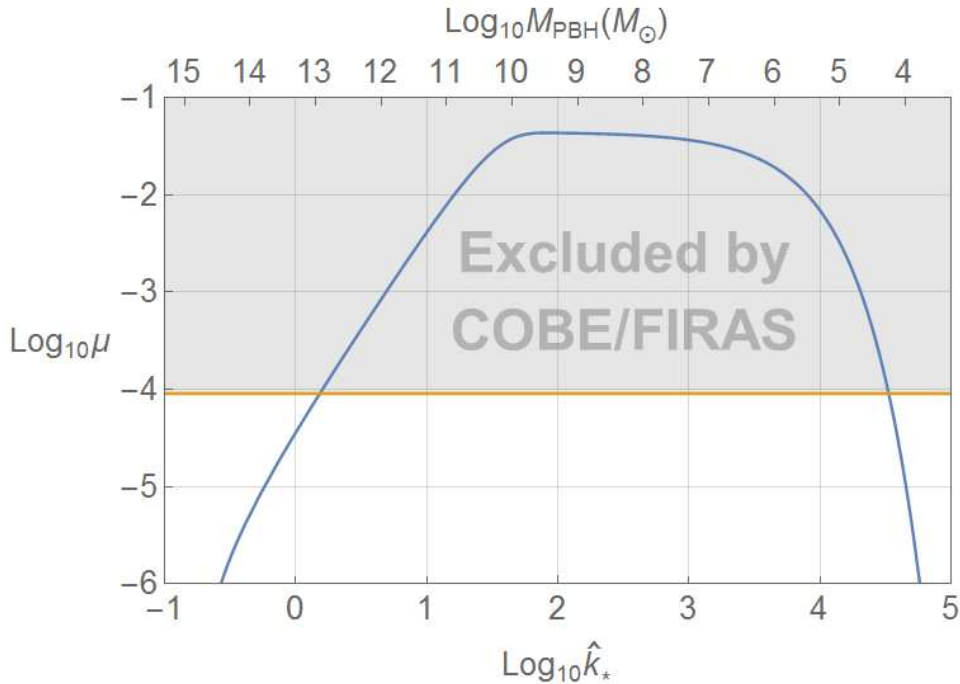}
\end{center}
\caption{
\bl{CMB} $\mu$ distortions generated from \bl{the delta-function-type power spectrum}, with $A_{\zeta}=0.02$. 
The horizontal line corresponds to the $2\sigma$ upper limit provided by 
COBE/FIRAS. \cat{This figure is a slightly modified version of Fig. 1 of
\cite{Kohri:2014lza}.
}
}
\label{mu}
\end{figure}

\bl{There is, however, a problem in explaining SMBHs by PBHs:} \red{simply} enhancing \soutb{the amplitude of the primordial 
perturbation} \fish{primordial perturbations} at \soutb{the PBH scale} \fish{suitable scales} to \soutb{the value that yields} \fish{yield a} sufficient amount of SMBHs, \bl{as stated above,}
is already excluded from the \cyan{observations of the energy spectrum of CMB photons} \cite{Carr:1993aq,Carr:1994ar,Chluba:2012we,Kohri:2014lza}.
\cyan{To see this, let us assume} Gaussianity of the primordial perturbation \fish{(non-Gaussian cases will be discussed later)}. \umi{Then, the} requirement that
produced PBHs are sufficient enough to explain the abundance of the observed SMBHs 
fixes the typical amplitude, or the root-mean-square amplitude, of the perturbations \cite{Carr:1994ar} \bl{to ${\cal O}(10^{-2})$}.
This amplitude is greater than the upper limit set by \bl{the} nondetection
of the distortion of the CMB spectrum by COBE \cite{Fixsen:1996nj};
this severely restricts the validity of the \red{scenario of PBHs \bl{whose initial mass exceeds} $\sim 10^4M_{\odot}-10^5M_{\odot}$} 
as the origin of the SMBHs, 
since these masses correspond to the shortest scales above which the dissipation of fluctuations causes CMB distortion\soutc{
, as is reviewed later}.
\cat{In \cite{Kohri:2014lza} this issue was revisited,
based on \cite{Chluba:2012we}, using the following delta-function-type spectrum of the curvature perturbation: 
\begin{equation}
P_\zeta(k)=2\pi^2A_\zeta k^{-2}\delta(k-k_*),
\end{equation}
and let us rewrite $k_*=\hat{k}_*\mathrm{Mpc}^{-1}$. 
Figure \ref{mu} shows a plot of CMB $\mu$ distortions resulting from this spike with $A_\zeta$ fixed to $0.02$, a value which is, roughly, necessary to produce a sufficient amount of PBHs assuming that primordial curvature perturbations are Gaussian. This figure is a slightly modified version of Fig. 1 of \cite{Kohri:2014lza}, and it shows that
any spike with $A_\zeta\gtrsim 0.02$ in a range 
$1 \lesssim {\hat k}_* \lesssim 3\times 10^4$ produces $\mu$ \dog{somewhat} larger than the
COBE/FIRAS upper bound.
Therefore, PBHs formed from a spike 
in the above \dog{range} of ${\hat k}_*$ are virtually excluded. 
This range of ${\hat k}_*$ can be translated into the PBH mass range as 
$2\times 10^4~M_\odot \lesssim M_{\rm PBH} \lesssim 2 \times 10^{13}~M_\odot$; that is, 
PBHs in this mass range are basically ruled out, at least for Gaussian perturbations.\footnote{\cat{
This point was also noted in \cite{Kawasaki:2012kn}, but they concluded PBHs with $M_{\rm{PBH}}>10^5M_\odot$ are severely constrained, 
and this upper bound of allowed masses is slightly larger than the one we obtain here ($M_{\rm{PBH}}\simeq 2\times 10^4M_\odot$). 
This is because, in \cite{Kawasaki:2012kn}, the upper bound was obtained by assuming only the perturbation modes which dissipate 
during the $\mu$ era, when dissipation of perturbations results in $\mu$ distortions efficiently, are severely constrained. Nevertheless, strictly speaking, since the transition to the $\mu$ era is gradual, 
the modes which dissipate before the onset of the $\mu$ era also cause $\mu$ distortions and, hence, are constrained, 
though relatively weakly.
This effect is taken into account in \cite{Kohri:2014lza}, based on \cite{Chluba:2012we}. }
}\soutc{, and hence we consider non-Gaussianity of primordial perturbations to evade this constraint on PBHs from CMB spectral distortions in 
this paper.}
}

\red{The r}oot of this constraint \bl{lies in} the fact that requiring the formation of \fish{a} sufficient
amount of PBHs inevitably leads to \red{\bl{relatively }large inhomogeneities} everywhere in the universe.
\bl{Even} though \soutb{the site of PBH formation is rare , where primordial 
perturbation is the order of unity,} \fish{PBH formation is extremely rare,}
\red{a} Gaussian \soutb{(or similar)} \red{probability density function (PDF)} \red{implies that perturbations everywhere else} \bl{are so large that their} 
\cyan{diffusion damping distorts the energy spectrum of CMB photons from a perfect Planck distribution (CMB distortion)} at a level excluded by COBE\soutb{ \bl{observations}
(see \red{the} left panel of Fig.~\ref{fig:pdf.eps})}.
\bl{Admittedly, there} is a possibility that PBHs whose initial mass is $\sim 10^4M_\odot-10^5M_{\odot}$\footnote{
Smaller PBHs are also potentially excluded by compact dark matter halos \cite{Kohri:2014lza} 
and acoustic reheating \cite{Jeong:2014gna,Nakama:2014vla}.
} grow to explain SMBHs of $10^{9}M_\odot-10^{10}M_\odot$ at high redshifts, 
as is argued in \cite{Kawasaki:2012kn},
but whether PBHs can grow to these masses is uncertain. One of the benefits of resorting to PBHs
is that one can create sufficiently large black holes in the early Universe due to collapse of primordial perturbations, 
but this benefit seems to have been partially lost due to CMB $\mu$ distortion.
Also, future experiments may reveal even more massive SMBHs at higher redshifts. 

\begin{figure}[tb]
  \begin{center}
    \includegraphics[width=130mm
    ]{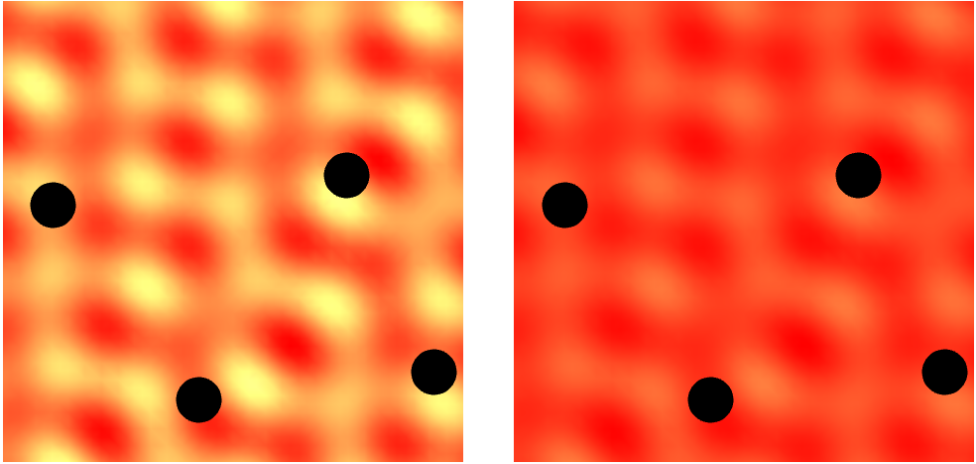}
  \end{center}
  \caption{\fish{An illustration of situations discussed in this paper. The black regions correspond to those where curvature perturbations become large during inflation and collapse to PBHs later during the radiation-dominated era. Normally, if a sufficient amount of PBHs is realized to explain the SMBHs, fluctuations whose wavelengths correspond to the mass of those PBHs are relatively large, as depicted in the \sea{left} panel, and, hence, they dissipate to produce CMB distortions larger than observational upper limits set by COBE. 
  In this paper, in order to explain SMBHs by PBHs \sea{without contradicting} this CMB distortion constraint, we discuss phenomenological inflation models that realize a sufficient probability of PBH formation to explain the SMBHs, while keeping fluctuations \sea{with corresponding wavelengths sufficiently small} outside these patches, as depicted in the \sea{right} panel, thereby
  evading the CMB distortion constraint. 
  }
}
  \label{real}
\end{figure}

In this light, we propose a novel inflationary scenario in which density perturbation\red{s} are generated
yielding PBHs \bl{whose initial mass is} \textit{larger} than $10^4M_\odot-10^5M_\odot$ as the origin of SMBHs \red{while evading}
the constraint \red{from CMB distortion mentioned} above. \fish{This can be accomplished by realizing a tiny fraction of patches where curvature perturbations become large during inflation, collapsing to PBHs later during the radiation-dominated era, while keeping the spectrum of curvature perturbations almost scale invariant outside those patches, as depicted in Fig. \ref{real}. Then, 
fluctuations whose wavelengths correspond to the masses of these PBHs, as the seeds of the SMBHs, are sufficiently small and, hence, the CMB distortion constraint can be evaded. }
\soutc{After revisiting CMB distortion constraints on PBHs in the next section, we}We will discuss a mechanism of how such 
\red{a} \soutb{non-Gaussian PDF} \fish{situation} can be realized in the framework of inflation\fish{, and then provide two toy models}.
\soutb{We also provide \cyan{two} simple toy model\cyan{s} in which such a mechanism is realized. }
We focus on the most massive SMBHs ($10^9M_\odot-10^{10}M_\odot$) observed \bl{at} high redshifts, 
\umi
{for which no compelling astrophysical explanations exist at the moment.}
In the last section, we discuss consequences of our scenario and
how it can be tested and distinguished fro\red{m a}strophysical explanations. 

\cyan{As already mentioned, simply preparing Gaussian perturbations whose dispersion is sufficiently large to 
generate PBHs as the seeds of SMBHs \bl{contradicts with} constraints on CMB distortion. 
One may first try to evade this by a monotonically decreasing PDF whose tail is considerably enhanced \bl{in comparison to that of a Gaussian PDF with the same dispersion}\soutb{, different from 
a bimodal PDF discussed in this paper}. In \soutb{the} Appendix \sora{A}, this possibility is briefly explored by calculating \bl{CMB spectral distortions} for a class of phenomenological models of PDFs. 
It turns out that 
it \bl{also works} (if such a PDF can \bl{indeed be} realized in some inflationary model, which we do not discuss \bl{in this paper}), 
but the PDF has to be \bl{hugely \soutb{different}} \fish{deviated} from a Gaussian PDF. 
}

\soutc{This paper is organized as follows:
In \S 2, constraints on the abundance of PBHs obtained from CMB $\mu$-\bl{distortions} are revisited. 
Then in}\cat{In the next section} \soutc{\S 3} we discuss inflationary models, in which PBHs can be produced whose mass and abundance 
are adjustable, in order to explain \bl{the} SMBHs observed at high redshifts, \fish{while evading CMB distortion constraints,} and 
\cat{then} we summarize and conclude in \S \cat{3}. 
\section{Supermassive black holes \red{formed} \bl{by collapse of} \\inflation\red{ary perturbations}}
\subsection{Basic idea}

\bl{Our} observable Universe consists of many small
patches \red{which become} causally disconnected during inflation.
For instance, if we consider a patch o\red{f c}omoving wave number $k$, 
it \cyan{becomes} decoupled from the other patches of the same size at a time when $k=aH$.
After this time, each patch evolves independently as if they themselves 
were \red{an} individual \red{Friedmann-Lemaitre-Robertson-Walker (FLRW)} universe.
If the inflation is caused by \bl{a single slowly rolling scalar} field,
only adiabatic perturbation\bl{s are} generated.
In this case, each patch follows the same trajectory in field space
and the difference between the patches is just \red{the} difference \bl{in the moment} when
the field value \bl{in each} takes a particular value.
On the other hand, if inflation is caused by multiple fields,
isocurvature \bl{perturbations} are also generated \sora{besides the adiabatic mode}.
Because of the presence of the \sora{former},
each patch follow\red{s a d}ifferent trajector\sora{y} in field space in general\fish{, and in the following} \soutb{.
In the following,} we assume \sora{such a situation}.
\begin{figure}[tb]
  \begin{center}
    \includegraphics[width=140mm,height=60mm]{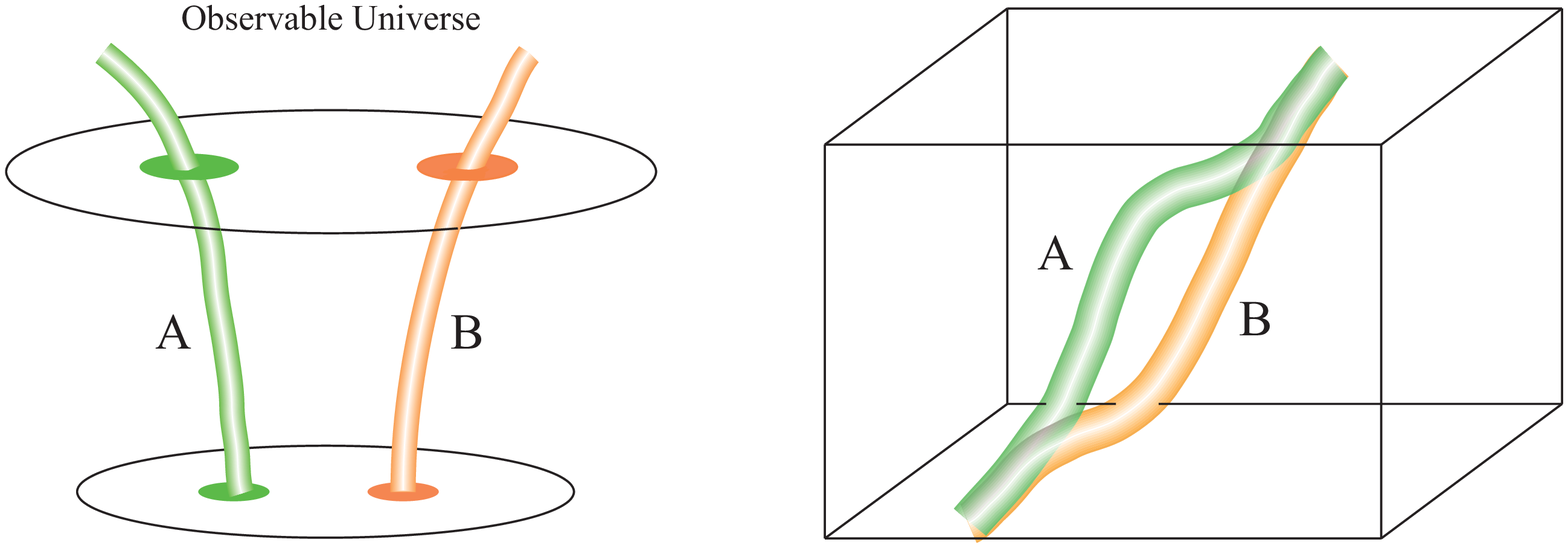}
  \end{center}
  \caption{Left figure:~This shows the separate universe picture
  \red{in which} patch\sora{es} A and B evolve independently as if each were \red{an} FLRW \red{u}niverse.
  Right figure:~Trajectories in field space corresponding to patches A and B, respectively.}
  \label{fig:trajectory.eps}
\end{figure}
 
Now, suppose that there are essentially only two different trajectories that each
patch can follow (see \red{the} \bl{right} panel of Fig.~\ref{fig:trajectory.eps}).
Let us label each trajectory by A and B, 
respectively (see \red{the} right panel of Fig.~\ref{fig:trajectory.eps}).
In general, \red{the} patch\red{es} corresponding to A and \red{the} patch\red{es} corresponding to B, 
after being causally disconnected, 
expand by \bl{a} different amount, namely, $N_A \neq N_B$ ($N_A\red{(}N_B\red{)}$ is the 
number of $e$-folds in \red{the} patch\red{es} A\red{(}B\red{)}, \bl{see the left panel of Fig.\ref{fig:trajectory.eps}}).
According to the $\delta N$ formalism \cite{Starobinsky:1986fxa,Salopek:1990jq,Sasaki:1995aw,Sasaki:1998ug,Wands:2000dp,Lyth:2004gb},
\fish{the difference in the number of $e$-folds} is equal to the curvature perturbation $\zeta$ o\red{n c}onstant
density hypersurface\red{s}.

It is known that if the region of interest has $\zeta$ exceeding $\zeta_c\simeq 1$,
such a region undergoes gravitational collapse to form a black hole when it reenters 
the Hubble horizon \cite{Carr:1974nx}.
\red{The t}hreshold value $\zeta_c$ depends on the perturbation
profile; there is a lot of literature in which \red{the} determination of 
$\zeta_c$ as well as its dependence on the perturbation profile
\bl{have} been investigated.
For instance, Shibata and Sasaki \cite{Shibata:1999zs} found that \cyan{$\zeta_c$ depends} on
the initial curvature profile and it \red{varies} at least in the range $(0.7,1.2)$ \red{(see also \cite{Niemeyer:1999ak,Hawke:2002rf,Polnarev:2006aa,Polnarev:2012bi,Harada:2013epa,Nakama:2013ica,Nakama:2014fra,Harada:2015yda})}. 
\red{However,} precise knowledge of $\zeta_c$ is not crucial for our discussions here \red{and so}
we simply take $\zeta_c=1$.

Let us assume that most of the patches
followed the trajectory A and the trajectory B is followed by only a \bl{tiny}
number of patches and that $\fish{N_B-N_A}>\zeta_c=1$.
Then, \red{the} patch\red{es} corresponding to B \red{distribute} sparsely, with each surrounded by 
patches corresponding to A, and each patch B has \bl{a} positive curvature perturbation $\fish{N_B-N_A}$.
In other words, large curvature perturbation\bl{s} of $\zeta > \zeta_c$ \bl{are} generated only in the patches B
and no \red{substantial} curvature perturbation is generated by the present mechanism in the patches A 
occupying most part of the universe. 
Because of our assumption that $\fish{N_B-N_A} > \zeta_c$, each patch B
turns into \red{a} BH upon horizon reentry.
Noting that \red{the} mass of the resultant BH is directly related to the comoving size of the patch\fish{es} B,
the time when the trajectories A and B start to deviate determines the BH mass.
\red{In this paper} we \red{consider two inflation models that can realize these situations with appropriately chosen model parameters.} 

Let us denote by $\beta$ the probability that \red{a} region \cyan{whose size is the same as that of patches B} collapses to \red{a} BH, namely, 
\begin{equation}
\beta = \frac{\rm number~of~patches~B}{\rm number~of~patches~A}.\label{ratio}
\end{equation}
\red{The rareness} of \red{the} patches B means $\beta \ll 1$\red{, 
which is required by observations a}s we will show below.

Observations of SMBHs at high redshifts suggest that one SMBH of $M_{\rm BH} \sim 10^{10} M_\odot$
exists roughly in every comoving volume $V$ of $1~{\rm Gpc}^3$ \cite{Fan:2003wd}.
Taking these \bl{numbers as fiducial values}, we find \red{the present energy density of these SMBHs normalized by the present critical density $\rho_c$,} \red{denoted by} $\Omega_{\rm BH,0}$\red{,} \umi{is} given by
\begin{equation}
\Omega_{\rm BH,0} = \frac{M_{\rm BH}}{\rho_c V} \approx 7 \times 10^{-11} 
\left( \frac{M_{\rm BH}}{10^{10} M_\odot} \right)
{\left( \frac{V}{{\rm Gpc}^3} \right)}^{-1}.
\end{equation}
In order to relate $\beta$ with $\Omega_{\rm BH,0}$, 
let us note that \red{the mass of a BH that formed at a} redshift $z$ is given by
\begin{equation}
M_{\rm BH} \simeq \frac{1}{2GH(z)},
\end{equation}
where $H(z)$ is the Hubble parameter at $z$.
From this equation, we find $M_{\rm BH} =6\times 10^{17}~M_\odot$ \red{if it is formed at} the matter-radiation equality $z=z_{\rm eq}$\red{.}
Hence BHs with $M_{\rm BH}\:\red{\lesssim}\: 10^{10} M_\odot$, which we are interested in,
formed in the \cyan{radiation-dominated} epoch.
Using $H(z) =H_0 {(1+z)}^2 \sqrt{\Omega_{r,0}}$\red{,} valid for $z > z_{\rm eq}$,
we have
\begin{equation}
1+z=2 \times 10^7 {\left( \frac{M_{\rm BH}}{10^{10}M_\odot} \right)}^{-1/2}.
\end{equation}
Then, using a relation $\Omega_{\rm BH,0}=\beta \Omega_{r,0} (1+z)$, we have
\begin{equation}
\beta = 4\times 10^{-14} {\left( \frac{M_{\rm BH}}{10^{10}M_\odot} \right)}^{3/2}
{\left( \frac{V}{{\rm Gpc}^3} \right)}^{-1}.\label{beta}
\end{equation}
Thus, observations require $\beta \ll 1$. 
\red{Note that the initial mass of PBHs does not have to be $\sim 10^{10}M_\odot$ to explain the observed SMBHs at high redshifts, 
since the mass of PBHs should grow to some extent, mainly after the matter-radiation equality \bl{with} the growth during radiation domination known to be quite limited. 
The accurate description of the growth of mass on a cosmological time scale would be a formidable task, 
which is beyond the scope of this work. 
However, we can adjust the typical mass of PBHs formed in our models simply by changing $\phi_{\rm{BH}}$ introduced later, 
so this issue \umi{does not} affect the feasibility of our model.
Also, it would be more natural to expect that only a fraction of SMBHs 
are bright enough to be observed at high redshifts and so the total number density of SMBHs, including those which are too dim to be observed,  would be larger than $\sim 1\mathrm{Gpc}^{-3}$ 
mentioned above. 
However, the uncertainty of $\beta$ stemming from these two issues 
does not affect the feasibility of our model, 
since $\beta$ turns out to only slightly affect $\bar{\chi}$, which is estimated later in (\ref{chibar}). 
}
 
\subsection{\cyan{Simple model 1: A hill on top of the $\phi^2$ potential}}
\begin{figure}[tb]
  \begin{center}
    \includegraphics[width=140mm,height=80mm]{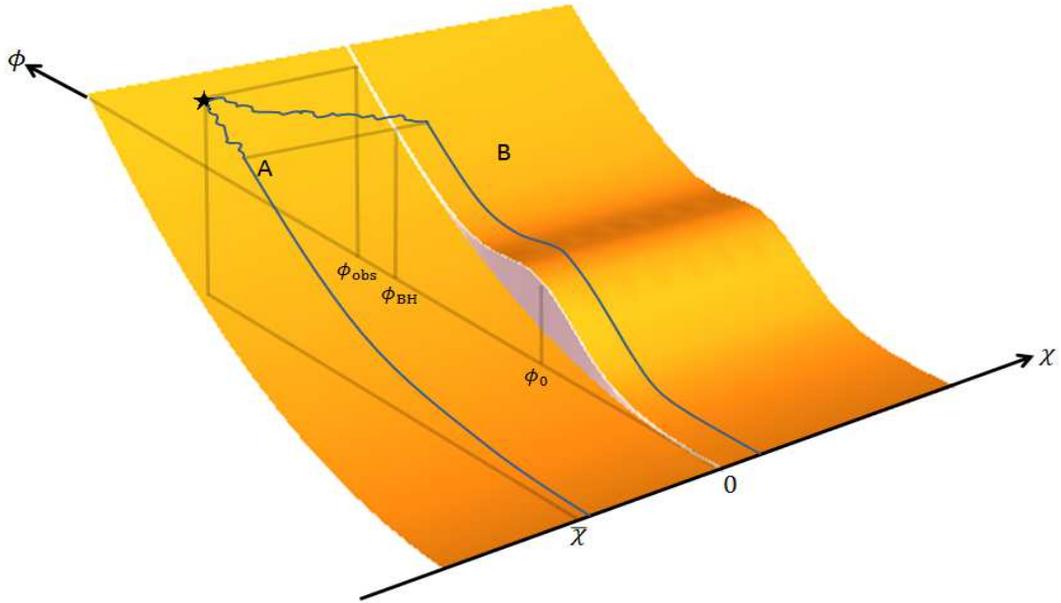}
  \end{center}
  \caption{\cyan{Illustrations of t}rajectories of mini universes A and B in field space for
  the potential given by Eq.~(\ref{model-potential}). 
  \cyan{
  Suppose there exists a hill at $\phi=\phi_0$ for $0<\chi$, and 
  all trajectories are assumed to start at $(\phi,\chi)\simeq (\phi_{\mathrm{obs}},\bar{\chi})$, denoted by the star in this figure. 
  \bl{The trajectories are zigzag \bl{for $\phi>\phi_{\mathrm{BH}}$} reflecting quantum fluctuation of $\chi$, while they are smooth for $\phi<\phi_{\mathrm{BH}}$ reflecting the classical nature of the time evolution.}
  If the absolute magnitude of $\bar{\chi}(<0)$ is sufficiently large,  
  only an extremely rare fraction of the patches of $\sim k_{\mathrm{BH}}$ enter into the region $0<\chi$, subsequently reaching the hill. 
  The amount of expansion is different between these two types of trajectories, and so 
  patches experiencing the hill are where the curvature perturbation is locally large. 
  If the hill is sufficiently wide and high, the amplitude of this curvature perturbation becomes order unity, 
  leading to the formation of PBHs. The mass and abundance of PBHs can be roughly controlled by the position of the hill and $\bar{\chi}$, 
  to explain SMBHs observed at high redshifts. 
  }}
  \label{fig:bump.eps}
\end{figure}

In this subsection, we provide a two-field inflation model in which 
\red{P}BHs as the observed SMBHs are produced by the mechanism we explained
in the previous subsection.
The Lagrangian density we consider is given by
\begin{equation}
{\cal L}=-\frac{1}{2} {(\partial \phi)}^2-\frac{1}{2} {(\partial \chi)}^2-
V(\phi) \left( 1+\theta (\chi) v(\phi) \right)\red{,}\label{lag}
\end{equation}
where $\theta(\chi)$ is the unit step function\footnote{
\bl{Strictly speaking the unit step function is 
unrealistic, but qualitatively the results of this paper are not affected as long as the transition at $\chi=0$ is sufficiently sharp.}
}.
Inflation is caused by the potential $V(\phi)$ for $\chi<0$ and 
$V(\phi) (1+v(\phi))$ for $\chi>0$.
To be definite, we adopt the following functions for $V(\phi)$ and $v(\phi)$:
\begin{equation}
V(\phi)=\frac{1}{2}m^2 \phi^2,~~~~~v(\phi)=\alpha \exp \left( -\frac{{(\phi-\phi_0)}^2}{2\mu^2} \right).
\label{model-potential}
\end{equation}
Here $\alpha$ is a positive dimensionless parameter.
Then, the field $\phi$ in the positive-$\chi$ region rolls \red{down} the potential
which is \bl{slightly} higher than \bl{that in} the negative-$\chi$ region.
Thus, trajector\red{ies} in \red{the} positive-$\chi$ region experienc\red{e a greater} number of $e$-folds 
than \red{those} in \red{the negative}-$\chi$ region.
In terms of the definition introduced previously,
trajector\red{ies} with negative/positive $\chi$ correspon\red{d t}o patch\red{es} A and B, 
respectively (see Fig.~\ref{fig:bump.eps}).

\bl{How} can the above inflation model realize the mechanism described in the previous subsection\bl{?}
\bl{To see this let us first} evaluate the initial condition of each patch 
of the comoving size $k_{\rm BH}^{\red{-1}}$ corresponding to the mass of SMBHs when each patch becomes
causally disconnected.
\bl{We} denote by $\cyan{\phi_{\mathrm{obs}}}$ and ${\bar \chi}$ the values of the scalar fields when 
the \red{current} observable universe crosses the Hubble horizon during inflation.
At this moment, all the patches of the comoving size corresponding to the SMBHs
are well deep inside the Hubble horizon an\umi{d t}ake the same values $(\cyan{\phi_{\mathrm{obs}}},{\bar \chi})$.
We require ${\bar \chi} <0$ so that the most region\red{s} of the universe \red{follow} \red{trajectories}
with negative $\chi$ afterwards.
By the time when $k_{\rm BH}$ becomes equal to $aH$, regions of comoving size \red{larger}
than $k_{\rm BH}^{-1}$ but smaller than $k_{\rm obs}^{-1}$\red{, the comoving scale of the current observable universe,}  have undergone classical
slow-roll motion associated with stochastic motion originating from redshifting 
of the short-wavelength vacuum fluctuations \cite{Starobinsky:1986fx}.
Thus, at the moment when $k_{\rm BH}=aH$, each patch of the comoving size $k_{\rm BH}^{-1}$
has randomly different field values centered at the values determined by
the classical slow-roll equations of motion.
\red{The d}istribution of the $\red{\chi}$ field value around the center\red{, in this case $\bar{\chi}$,} is approximately Gaussian and its variance
is given by \cite{Starobinsky:1982ee}
\begin{equation}
\langle {(\chi-{\bar \chi})}^2 \rangle \simeq \frac{H^2}{4\pi^2} (N_{\rm obs}-N_{\rm BH}),
\label{chi-distribution}
\end{equation}
where $N_{\rm obs}-N_{\rm BH}$ is the number of
$e$-folds between the time when the observable universe crossed the Hubble horizon and 
the one when the size of $k_{\rm BH}^{-1}$ crossed the Hubble horizon.
Approximating that $H$ remains almost constant during that period, we have
$N_{\rm obs}-N_{\rm BH} \simeq \ln ( \red{k_{\rm BH}/k_{\rm obs}})$.

After the time $k_{\rm BH}=aH$, each patch of the comoving size $k_{\rm BH}^{-1}$ becomes
causally disconnected and the fields on each patch evolve independently from the others.
Adopting the viewpoint of the separate universe picture \cite{Wands:2000dp}, we make an assumption
that the fields on each patch behave as spatially uniform fields which obey classical 
equations of motion for the homogeneous fields in the FLRW spacetime, whose 
expansion is also determined by the field values in the same patch. 
Each patch follow\red{s d}ifferent trajector\red{ies} in field space due to different
field values at the time $k_{\rm BH}=aH$.
However, because of the special form of the potential \red{we consider}, 
only whether $\chi$ is positive or negative matters in terms of the number of $e$-folds.
In this sense, there are \red{essentially} only two trajectories in field space (trajector\red{ies} with positive $\chi$
and negative $\chi$) and the model can effectively realize the mechanism
described previously.
The condition that the patch\red{es} \bl{with} positive $\chi$ (\red{the} patch\red{es} B in the language introduced previously)
\bl{have curvature perturbations} greater than $\zeta_c$ imposes constraints among the model parameters.
In addition, 
in order for the above inflation model to \bl{successfully explain} the origin of SMBHs, 
the model also needs to achieve the observationally suggested values of $\beta$ and $M_{\rm BH}$, 
\bl{as discussed later.}

\ao{
\bl{We describe} how to calculate \bl{curvature perturbations \sora{on scale}} $k_{\mathrm{BH}}^{-1}$, 
working in a box of comoving size $k_{\rm{obs}}^{-1}\sim {\cal O}$(Gpc). 
Let us first discuss the amplitude of the curvature perturbation on $k_{\rm{BH}},$ when this mode exits the horizon at $t_{\rm{BH}}$. 
As is discussed shortly, the effects of the hill 
are chosen to be negligible up to $\phi_{\rm{BH}}$.
First, the amplitude of field fluctuations 
$\delta \phi$ and $\delta \chi$ on flat slices are given by (see e.g. \cite{Liddle-Lyth})
\begin{equation}
{\cal P}_{\delta \phi,\delta \chi}(t_{\rm{BH}},k_{\rm{BH}})=\left(\frac{H_{k_{\rm{BH}}}}{2\pi}\right)^2,
\end{equation}
where $H_{k_{\rm{BH}}}$ is the Hubble parameter when the mode $k_{
\rm{BH}}$ exits the horizon. 
We assume that the energy density of $\chi$ is always negligible, and so the curvature perturbation $\zeta$ on uniform-density slices at $t_{\rm{BH}}$
is solely determined by $\delta \phi$ and is given by 
$\zeta=-H\delta\phi/\dot{\phi}$. Hence, the power spectrum of the curvature perturbation at $t_{\rm{BH}}$ is 
\begin{equation}
{\cal P}_\zeta(t_{\rm{BH}},k_{\rm{BH}})=\frac{1}{4\pi^2}\left(\frac{H^2}{\dot{\phi}}\right)^2
=\frac{1}{24\pi^2M_{\rm{Pl}}^4}\frac{V}{\epsilon}.\label{spectrum}
\end{equation}
Without the presence of the hill ($\alpha=0$), 
fluctuations on $k_{\rm{BH}}$ just correspond to the 
time difference on the essentially same trajectory, 
noting that in this case $\chi$ does not affect cosmic expansion and hence plays no role, 
and the curvature perturbation is conserved after $k_{\rm{BH}}$ exits the horizon. 
Also, perturbations in this case are Gaussian and almost scale invariant. These perturbations are determined by $V(\phi)$ and we choose it so that ${\cal P}_\zeta\sim {\cal O}(10^{-9})$ to match observations on large scales. 
In this case, the probability of PBH formation is vanishingly small. Next, let us consider the effects of the hill ($\alpha\neq 0$). 
After $k_{\rm{BH}}^{-1}$ exits the horizon, each region of $k_{\rm{BH}}^{-1}$ can be regarded as evolving as an independent FLRW universe \cite{Wands:2000dp}. The metric on uniform density slices may be written as $d\sora{s}^2=-dt^2+\tilde{a}\sora{^2}(t,\vec{x})d^{\sora{2}}\vec{x}$, 
where $\tilde{a}(t,\vec{x})=a(t)\exp[\zeta(t,\vec{x})]$ is the local scale factor, $a(t)$ is the global scale factor and $\zeta(t,\vec{x})$ is the curvature perturbation. 
Here and hereafter, the position-dependent quantities are understood to be those smoothed over the comoving scale of $k_{\rm{BH}}^{-1}$, not over the 
Hubble radius at each moment. 
Let us consider two patches A and B of $k_{\rm{BH}}^{-1}$ around points $\vec{x}_A$ and $\vec{x}_B$, and assume that 
in most of the regions inside the patch A(B) $\chi$ continues to be negative (positive) for $t>t_{\rm{BH}}$.  
Note that, even if $\chi(t_{\rm{BH}},\vec{x}_B)>0$, this does not 
ensure the positivity of $\chi$ in most of the regions inside the patch B 
for $t>t_{\rm{BH}}$. To see this first recall that, after $t_{\rm{BH}}$, the field values $\phi$ and $\chi$ smoothed over the Hubble radius at 
each point keep randomly fluctuating by $\sim H$ over the time scale $\sim H^{-1}$. 
This means that, naively, if $\chi(t_{\rm{BH}},\vec{x}_B)>0$ but 
$\chi(t_{\rm{BH}},\vec{x}_B)\ll H,$ 
roughly half of the region in the patch B would end up having $\chi<0$; 
more precisely, due to the sharp wall at $\chi=0$, hindering crossing from $\chi<0$ to $\chi>0$ for $t>t_{\rm{BH}}$, 
actually more than half of the region in the patch B would end up having $\chi<0$. 
Hence, we need $\chi(t_{\rm{BH}},\vec{x}_B)>{\cal O}(1)H$
to ensure the positivity of $\chi$ in most of the regions in the patch B for $t>t_{\rm{BH}}$. 
The curvature perturbation at $(t_{\rm{BH}},\vec{x}_{A,B})$, $\zeta(t_{\rm{BH}},\vec{x}_{A,B})$, is of order ${\cal O}(10^{-5})$, 
the same as the case without the hill as explained above since the effects of the hill are negligible up to $t_{\rm{BH}}$. 
When the hill is present, the inflaton trajectories for $t>t_{\rm{BH}}$ qualitatively differ depending on $\chi$; in this case, the curvature perturbation of the patch B grows for $t_{\rm{BH}}<t<t_{\rm{end}}$, where $t_{\rm{end}}$ corresponds to the end of the inflation, 
and this growth is entirely determined by the difference in the overall expansion histories of the patches A and B for $t>t_{\rm{BH}}.$ 
This is because, as long as $\chi$ stays negative (positive) in most regions in the patch A (B), 
quantum fluctuations of $\chi$ on the Hubble radius arising after $t_{\rm{BH}}$ essentially do not play any role, 
in the sense that it no longer affects expansion. 
Also, quantum fluctuations of $\phi$ on the Hubble radius arising after $t_{\rm{BH}}$ keep being converted to curvature perturbation\bl{s} on $k>k_{\rm{BH}}$, 
but this does not affect the curvature perturbation on $k_{\rm{BH}}^{-1}$ either. 
To calculate the growth of the curvature perturbation for $t>t_{\rm{BH}}$ due to the difference in the expansions, 
let us define the local Hubble parameter $H(t,\vec{x})$ by 
\begin{equation}
H(t,\vec{x})\equiv \frac{\dot{\tilde{a}}(t,\vec{x})}{\tilde{a}(t,\vec{x})}=\frac{\dot{a}(t)}{a(t)}+\dot{\zeta}(t,\vec{x}).\label{local}
\end{equation}
\bl{Let us temporarily adopt the slow-roll approximations to illustrate how to evaluate curvature perturbations, though we use exact equations later.}
During inflation, the equation of motion for $\phi$ in the patch A is given by
\begin{equation}
3H {\dot \phi}+V'(\phi) \simeq 0,~~~~~
H^2 \simeq \frac{1}{3M_{\rm Pl}^2}V(\phi),\label{slowroll}
\end{equation}
where a prime denotes differentiation with respect to $\phi$ and $M_{\mathrm{Pl}}$ is the reduced Planck mass,
and for the patch B
\begin{equation}
3H {\dot \phi}+\big[ V(\phi) (1+v(\phi)) \big]' \simeq 0,~~~~~
H^2 \simeq \frac{1}{3M_{\rm Pl}^2}V(\phi) (1+v(\phi)).
\end{equation}
The numbers of $e$-folds of the patches A and B from $t_{\rm{BH}}$ to $t_{\rm{end}}$ are given by
\begin{equation}
N_A=\frac{1}{M_{\rm Pl}^2} \int_{\phi_{\rm end}}^{\phi_{\rm BH}} d\phi~
\frac{V(\phi)}{V'(\phi)},~~~~~
N_B=\frac{1}{M_{\rm Pl}^2} \int_{\phi_{\rm end}}^{\phi_{\rm BH}} d\phi~
\frac{V(\phi) (1+v(\phi))}{\big[ V(\phi) (1+v(\phi))\big]\red{'}}.
\end{equation}
PBH formation is determined by the difference in the curvature perturbation at the end of inflation, 
since thereafter it is conserved
, 
and from (\ref{local}) it is expressed as
\begin{equation}
\zeta(t_{\rm{end}},\vec{x}_B)-\zeta(t_{\rm{end}},\vec{x}_A)
=\zeta(t_{\rm{BH}},\vec{x}_B)-\zeta(t_{\rm{BH}},\vec{x}_A)+\Delta N,\quad \Delta N\equiv \fish{N_B-N_A}.
\end{equation}
As mentioned above, $\zeta(t_{\rm{BH}},\vec{x}_{A,B})\sim {\cal O}(10^{-5})$, while we are interested in situations where $\Delta N\sim 1$ to produce PBHs, 
and so we can safely neglect $\zeta(t_{\rm{BH}},\vec{x}_{A,B})$ and focus on $\Delta N$ in the following. 
}

\cyan{
Let us calculate the relationship between $\phi_{\rm{BH}}$ and $M_{\rm{BH}}$.
The mass of PBHs $M_{\rm{BH}}$ is roughly estimated by the horizon mass at the moment when the comoving scale $k_{\rm{BH}}$ reenters the horizon, 
from which one finds
\begin{equation}
M_{\rm{BH}}\sim 2.2\times 10^{13}M_\odot\left(\frac{k_{\rm{BH}}}{1\rm{Mpc}^{-1}}\right)^{-2}.
\end{equation}
This can be inverted as follows:
\begin{equation}
k_{\rm{BH}}\sim47\left(\frac{M_{\rm{BH}}}{10^{10}M_\odot}\right)^{-1/2}\rm{Mpc}^{-1}.\label{kbh}
\end{equation}
Noting the following relation
\begin{align}
&\ln\maru{\f{k_{\mathrm{BH}}}{k_{\rm{obs}}}}\simeq N_{\mathrm{obs}}-N_{\mathrm{BH}} =\int_{t_{\mathrm{obs}}}^{t_{\rm{BH}}}dtH
\simeq \f{1}{M_{\mathrm{Pl}}^2}\int_{\phi_{\mathrm{BH}}}^{\phi_{\mathrm{obs}}}d\phi\f{V}{V'}
=\f{1}{4M_{\mathrm{Pl}}^2}(\phi_{\mathrm{obs}}^2-\phi_{\mathrm{BH}}^2)\label{log}
\end{align}
and setting $k_{\rm{obs}}=1\rm{Gpc}^{-1}$, we obtain
\begin{equation}
\phi_{\rm{BH}}=
\sqrt{
 \phi_{\rm{obs}}^2-4M_{\rm{Pl}}^2\log\left(\frac{k_{\rm{\blue{BH}}}}{k_{\rm{\blue{obs}}}}\right)
}
\simeq 13\sqrt{1+0.01\log\left(\frac{M_{\rm{BH}}}{10^{10}M_\odot}\right)},
\end{equation}
\blue{where we have set $N_{\rm{obs}}=55\;(\phi_{\rm{obs}}\simeq 14.8M_{\rm{Pl}})$.}
Note that the dependence of $\phi_{\rm{BH}}$ on $M_{\rm{BH}}$ is very weak; for instance, if we set $M_{\rm{BH}}=1M_\odot$, $\phi_{\rm{BH}}\simeq 12M_{\rm{Pl}}$. 
Therefore, though we assume $M_{\rm{BH}}=10^{10}M_\odot$ and $\phi_{\rm{BH}}=13M_{\rm{Pl}}$ in the following, 
our analysis is valid for other masses as well. 
For instance, one may choose the typical initial mass of PBHs to be smaller than $10^{10}M_\odot$, taking into account possible mass growth of PBHs.
}

\cyan{\bl{F}or each $\mu$, \bl{the width of the hill,} $\phi_0$ has to be sufficiently smaller than $\phi_{\rm{BH}}$,}
\bl{so} that the sharp wall of the potential at $\chi=0$ 
does not prevent stochastic motion of $\chi$ from crossing the wall at $k_{\rm BH}=aH$.
The criterion that the stochastic motion can cross over the wall freely 
is that \red{the kinetic energy of $\chi$ field, $\sim H^4$}, is larger than the potential \cyan{gap} \red{at $\phi=\phi_{\mathrm{BH}}$}, \bl{otherwise} the wall blocks the stochastic motion effectively 
and $\chi$ \bl{cannot} enter \red{the} positive region.
\red{T}he height of the potential wall at the peak $\phi=\phi_0$ is given by $\alpha V(\phi_0)$ 
and this is much larger than $H^4$ in our present model
\red{for a range of $\alpha$ in which ${\cal O}(1)$ difference of the number of $e$-folds
arises between A and B.}
Thus, $\phi_{\rm BH}$ must be located \bl{sufficiently} far from the peak where 
the height of the wall is smaller than $H^4$,
\bl{and this requirement determines the position of the hill as follows. 
We introduce $R$ by}
\begin{equation}
R\equiv\frac{H^4}{V(\phi_{\rm{BH}})v(\phi_{\rm{BH}})}
\simeq\frac{m^2\phi_{\rm{BH}}^2}{18\alpha\blue{M_{\rm{Pl}}^4}\exp\{-(\phi_{\rm{BH}}-\phi_0)^2/2\mu^2\}}
\label{condition}
\end{equation}
and rewrite the exponential factor here by defining $\nu$ as $\phi_{\rm{BH}}=\phi_0+\nu\mu$, 
then we can solve for $\nu$ as
\begin{equation}
\bl{
\nu=
\left[2\ln\left(\frac{18\alpha R M_{\mathrm{Pl}}^4}{m^2\phi_{\mathrm{BH}}^2}\right)
\right]^{1/2}
\simeq 6.4\left(
1+0.05\left[\ln\left(\frac{\alpha}{0.06}\right)+\ln R\right]
\right)^{1/2},
}
\end{equation}
where we have set $\phi_{\rm{BH}}=13M_{\rm{Pl}}$ and $m=3\times 10^{-6}M_{\rm{Pl}}$. 
Hence in the following we fix $\nu=\bl{6.4}$. 
\sky{
Here we assume the crossing to the positive $\chi$ region happens only at $\phi=\phi_{\rm{BH}}=13M_{\rm{Pl}}$, 
leading to the monochromatic mass function of PBHs at $M_{\rm{BH}}\simeq 10^{10}M_\odot$. 
Strictly speaking however, the masses would be distributed around the mass determined by $\phi_{\mathrm{BH}}$, 
and this mass spectrum is determined by the following two effects. 
First, the crossing to the positive $\chi$ region can in principle also occur when $\phi>\phi_{\rm{BH}}$, 
though the probability of these cases is exponentially suppressed, since the probability of reaching $\chi=0$ becomes rapidly rarer as $\phi$ is increased. 
This effect determines the tail of the mass function at larger masses. 
Second, the crossing can occur even for $\phi<\phi_{\rm{BH}}$, though the probability would be increasingly suppressed as $\phi$ becomes closer to $\phi_0$ due to the gap of the potential at $\chi=0$; to quantify this effect, the probability of $\chi$ jumping over the gap by stochastic motion has to be calculated. 
This issue is explored in \soutb{the} Appendix B.
}
\color{black}
If the hill, described by $v(\phi)$, is sufficiently high, the slow-roll conditions are violated near the hill located around $\phi_0<\phi_{\mathrm{BH}}$. 
Hence, we use the following equations
without the slow-roll approximations, 
to solve for the time evolution of $\phi$ for $\phi<\phi_{\mathrm{BH}}$:
\begin{equation}
\ddot{\phi}+3H\dot{\phi}+[V(\phi)(1+v(\phi))]'=0,
\end{equation}
\begin{equation}
H^2=\frac{1}{3M_{\mathrm{Pl}}^2}\left(\frac{\dot{\phi}^2}{2}+V(\phi)(1+v(\phi))\right).
\end{equation}
It \umi{is} convenient to use the $e$-folds $N$ as the time variable \umi{defined as evolving backward in time}, 
then since 
\begin{equation}
N=\int_t^{t_{\mathrm{end}}}Hdt\rightarrow \frac{\partial}{\partial t}=-H\frac{\partial}{\partial N},
\end{equation}
the above can be rewritten as
\begin{equation}
\phi_{NN}+\left(\frac{H_N(\phi,\phi_N,\phi_{NN})}{H(\phi,\phi_N)}\blue{-3}\right)\phi_N+\frac{1}{H^2}[V(\phi)(1+v(\phi))]'=0,
\end{equation}
\begin{equation}
H^2(\phi,\phi_N)=\frac{V(\phi)[1+v(\phi)]/3M_{\mathrm{Pl}}^2}{1-{\phi_N}^2/6M_{\mathrm{Pl}}^2},
\end{equation}
where the subscripts $N$ denote differentiation with respect to $N$. 

\begin{figure}[t]
\begin{center}
\includegraphics[width=13cm,keepaspectratio,clip]{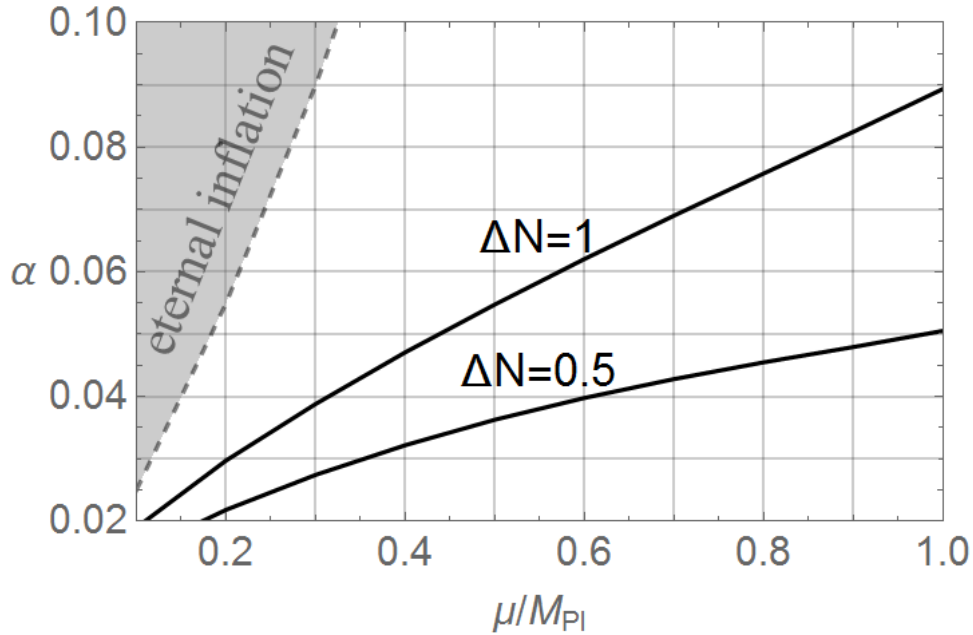}
\end{center}
\caption{
\bl{
A contour plot of $\Delta N$ caused by a hill on top of the 
$\phi^2$ potential. The shaded region corresponds to parameters leading to eternal inflation at the patches B.
}
}
\label{phisquare}
\end{figure}

We use these exact equations only for $\phi<\phi_{\rm{BH}}$, where 
slow-roll conditions may be violated. 
The initial conditions to solve the above exact equations are provided at $\phi_{\mathrm{BH}}$ using the slow-roll approximations, always valid for $\phi>\phi_{\rm{BH}}$, as follows.
For the case of the $\phi^2$ potential, we have
\begin{equation}
\umi{
N=\frac{1}{4M_{\mathrm{Pl}}^2}(\phi^2-\phi_{\mathrm{end}}^2)
}
\end{equation}
and its differentiation with respect to $N$
\begin{equation}
1=\blue{\boldmath{+}}\phi\phi_N/2M_{\mathrm{Pl}}^2,
\end{equation}
so the initial conditions to solve the above equation of motion are 
\begin{equation}
N=N_{\mathrm{BH}}=\frac{1}{4M_{Pl}^2}(\phi_{\mathrm{BH}}^2-\phi_{\mathrm{\umi{end}}}^2),
\quad \phi=\phi_{\mathrm{BH}},
\quad \phi_N=\phi_{\mathrm{BH},N}=\frac{2M_{\mathrm{Pl}}^2}{\phi_{\mathrm{BH}}}.
\end{equation}
With these initial conditions, we solve the equation of motion up to $\phi_{\mathrm{end}}=\sqrt{2}M_{\mathrm{Pl}}$, corresponding to $\epsilon=1$ in the slow-roll approximation, for different parameters describing the hill $\alpha$ and $\mu$. The moment $N_{\mathrm{end,B}}$ when $\phi_{\mathrm{end}}$ is reached depends on the \umi{shape of the }hill at $0<\chi$, so $N_{\mathrm{end,B}}=N_{\mathrm{end,B}}(\alpha,\mu)$. 
Then the curvature perturbation is $\Delta N(\alpha,\mu)=\blue{N_{\mathrm{end,A}}-N_{\mathrm{end,B}}(\alpha,\mu)}$\footnote{
$\phi_{\mathrm{obs}}$ is chosen so that the $e$-folds in the A patches at $\phi=\phi_{\mathrm{end}}(=\sqrt{2}M_{\mathrm{Pl}}$ for the case of $\phi^2$ potential),
$N_{\mathrm{end},A}$, is zero in the slow-roll approximation, 
but $N_{\mathrm{end},A}$ deviates from zero with the numerical calculation without the slow-roll approximation, and the curvature perturbation should be defined as the deviation from that value. 
}.
A contour plot of $\Delta N$ is shown in Fig.\ref{phisquare}. 
\bl{For the same $\mu$, if $\alpha$ is larger, the hill is 
higher, and therefore $\Delta N$ is larger. For the same $\alpha$, 
if $\mu$ is smaller, $\phi_0$ is larger, and the height of the hill is higher, and hence $\Delta N$ is larger.}
If the hill is sufficiently high, 
$\phi_N$ becomes smaller than $H/2\pi$ near the hill, which means quantum motion is more important than classical motion, and hence eternal inflation occurs in the patches B. 
The parameter region leading to eternal inflation is also shown in Fig.\ref{phisquare}. 
\color{black}
To conclude, there is a parameter space where the curvature perturbation exceeds unity 
and, hence, PBHs can be formed; 
the resulting SMBHs have masses
around $10^{10}M_\odot$.

Finally, let us determine the initial value $\bar \chi$ of $\chi$ which leads to \red{an} observationally suggested value of $\beta$.
As already mentioned, $\chi$ on the patch\red{es} corresponding to $k_{\rm BH}$ \cyan{when $\phi=\phi_{\mathrm{BH}}$}
is randomly distributed around the central value $\bar \chi$ with its variance
given by Eq.~(\ref{chi-distribution}).
As a result, \cyan{noting (\ref{ratio})} $\beta$ is given by
\begin{equation}
\ao{
\beta \simeq \int_{0}^\infty d\chi~\frac{1}{\sqrt{2\pi} \sigma_\chi} 
\exp \left( -\frac{{(\chi-{\bar \chi})}^2}{2 \sigma_\chi^2} \right)
\simeq -\frac{\sigma_\chi}{\sqrt{2\pi} {\bar \chi}} 
\exp \left( -\frac{{\bar \chi}^2}{2 \sigma_\chi^2} \right),
}
\end{equation}
\bl{where we have used the fact} that the integral
picks up only the high-$\sigma$ tail of the Gaussian distribution (\bl{recall} that ${\bar \chi}<0$).
\sky{As explained previously, our calculation of the curvature perturbation 
at patches B is valid for $\chi(t_{\rm{BH}},\vec{x}_B)>{\cal O}(1)H$, 
so the lower bound of the integration here should be strictly speaking taken as 
${\cal O}(1)H$, but this only affects $\bar{\chi}$, evaluated below, only slightly.}
Solving the above equation for $\red{\bar{\chi}}$ yields
\begin{equation}
{\bar \chi}=-\sigma_\chi \sqrt{W_0 \left( \frac{1}{\ao{2}\pi \beta^2} \right)},
\end{equation}
where $W_0$ is the Lambert function.
Using the expansion of $W_0(x)$ for large $x$ given by 
$W_0(x)=\ln x-\ln \ln x+{\cal O}(1)$, we have
\begin{equation}
{\bar \chi} \simeq - \frac{H}{2\pi} \sqrt{N_{\rm obs}-N_{\rm BH}} 
\bigg[ -\ln (\ao{2}\pi \beta^2) -\ln( -\ln (\ao{2}\pi \beta^2) ) \bigg]^{\red{1/2}}.\label{chibar}
\end{equation}
\bl{That is}, \red{the} observed abundance of SMBHs can be realized if ${\bar \chi}$ takes
\bl{this} value.

\cyan{
\subsection{Simple model 2: A hill on top of the $R^2$-inflation\bl{-}type potential}
The $\phi^2$ potential considered in the previous subsection is somewhat disfavored by 
the Planck data \cite{Ade:2015lrj}. 
However, our mechanism can work for other types of potentials, 
including those favored by the Planck data. 
To see this\sora{,} in this subsection we consider a hill on top of the following potential: 
\begin{equation}
V(\phi)=\frac{3M^2M_{\mathrm{Pl}}^2}{4}
\nami{1-\exp\maru{-\sqrt{\f{2}{3}}\f{\phi}{M_{\mathrm{Pl}}}}}^2.\label{Starobinsky}
\end{equation}
This can be obtained by a conformal transformation (see, e.g.,\sora{\cite{Maeda:1987xf}}) of 
$R^2$ inflation \cite{Starobinsky:1980te}, which is so far favored by the Planck data. 
The parameter $M$ is fixed by the COBE-WMAP normalization
of the amplitude of the curvature perturbations as follows 
(see, e.g., \cite{Takeda:2014qma}):
\begin{equation}
M\simeq 10^{-5}M_{\mathrm{Pl}}\f{4\pi \sqrt{30}}{N_{\mathrm{obs}}}
\maru{\f{{\cal P}(k_*)}{2\times 10^{-9}}}^{1/2}
\simeq 1.25\times10^{-5}M_{\mathrm{Pl}}\maru{\f{N_{\mathrm{obs}}}{55}}^{-1}
\maru{\f{{\cal P}(k_*)}{2\times 10^{-9}}}^{1/2}.
\end{equation}
If we define $\phi_f$ by $\epsilon=1$, then $\phi_f=\sqrt{\f{3}{2}}\log\maru{1+\f{2}{\sqrt{3}}}M_{\mathrm{Pl}}\simeq 0.94M_{\mathrm{Pl}}.$
$N_{\mathrm{\umi{obs}}}$ for this model is given by 
\begin{equation}
N_{\mathrm{\umi{obs}}}=\f{3}{4}\nami{
 \exp\maru{\sqrt{\f{2}{3}}\f{\phi_{\mathrm{obs}}}{M_{\mathrm{Pl}}}}
 -\exp\maru{\sqrt{\f{2}{3}}\f{\phi_f}{M_{\mathrm{Pl}}}}
}-\f{\sqrt{6}}{4M_{\mathrm{Pl}}}(\phi_{\mathrm{obs}}-\phi_f).
\end{equation}
This can be approximately solved for $\phi_{\mathrm{obs}}$ (neglecting the last two terms above) as 
\begin{equation}
\phi_{\mathrm{obs}}\simeq\sqrt{\f{3}{2}}M_{\mathrm{Pl}}\log\nami{\f{1}{3}
\maru{4N_{\mathrm{\umi{obs}}}+2\sqrt{3}+3}}.
\end{equation}
We set $N_{\mathrm{\umi{obs}}}=55$ and then $\phi_{\mathrm{obs}}\simeq5.3M_{\mathrm{Pl}}.$
Then $\phi_{\mathrm{BH}}$ can be determined as follows:
\begin{align}
N_{\mathrm{obs}}-N_{\mathrm{BH}}&\simeq \ln\maru{\f{k_{\mathrm{BH}}}{k_{\mathrm{obs}}}}
\simeq\dog{11}\nami{1\dog{-0.1}\log_{10}\maru{\f{M_{\mathrm{BH}}}{10^{\dog{10}}M_\odot}}}\nonumber\\
&\simeq \f{3}{4}\nami{
 \exp\maru{\sqrt{\f{2}{3}}\f{\phi_{\mathrm{obs}}}{M_{\mathrm{Pl}}}}
 -\exp\maru{\sqrt{\f{2}{3}}\f{\phi_{\mathrm{BH}}}{M_{\mathrm{Pl}}}}
},
\end{align}
from which
\begin{equation}
\phi_{\mathrm{BH}}\simeq 5.0M_{\mathrm{Pl}}\kaku{
 1+0.56\log_{10}\nami{
  1\dog{+0.025}\log_{10}\maru{\f{M_{\mathrm{BH}}}{10^{10}M_{\odot}}}
 }
}.
\end{equation}
The \bl{ratio} corresponding to (\ref{condition}) is 
\begin{equation}
R\equiv \f{H^4}{V(\phi_{\mathrm{BH}})v(\phi_{\mathrm{BH}})}\simeq
\f{M^2}{12\alpha M_{\mathrm{Pl}}^2}\exp\nami{\f{(\phi_{\mathrm{BH}}-\phi_0)^2}{2\mu^2}}
\nami{1-\exp\maru{-\sqrt{\f{2}{3}}\f{\phi_{\mathrm{BH}}}{M_{\mathrm{Pl}}}}}^2.
\end{equation}
Once more, let us rewrite $\exp[(\phi_{\mathrm{BH}}-\phi_0)^2/2\mu^2]=\exp (\nu^2/2)$ and solve for $\nu$ to obtain
\bl{
\begin{align}
\nu&=\left(2\ln\left[\frac{12\alpha RM_{\mathrm{Pl}}^2}{M^2}\left\{1-\exp\left(-\sqrt{\frac{2}{3}}\frac{\phi_{\mathrm{BH}}}{M_{\mathrm{Pl}}}\right)\right\}^{-2}\right]\right)^{1/2}\nonumber\\
&\simeq 6.7 \left(1+0.045\left[\ln\left(\frac{\alpha}{0.06}\right)+\ln R\right]\right)^{1/2}.
\end{align}
The initial conditions to be provided at $\phi_{\mathrm{BH}}$ are
\begin{equation}
N=N_{\mathrm{BH}}=\frac{3}{4M_{Pl}^2}\left[\exp\left(\sqrt{\frac{2}{3}}\phi_{\mathrm{BH}}\right)-\exp\left(\sqrt{\frac{2}{3}}\phi_{\mathrm{\umi
{end}}}\right)\right],
\end{equation}
\begin{equation}
\phi=\phi_{\mathrm{BH}},
\quad \phi_N=\phi_{\mathrm{BH},N}=\frac{2\sqrt{2}M_{\mathrm{Pl}}^2}{\sqrt{3}}\exp\left(-\sqrt{\frac{2}{3}}\phi_{\mathrm{BH}}\right).
\end{equation}
}
A contour plot of $\Delta N$ in this case is shown in Fig. \ref{Rsquare}. 
\begin{figure}[t]
\begin{center}
\includegraphics[width=13cm,keepaspectratio,clip]{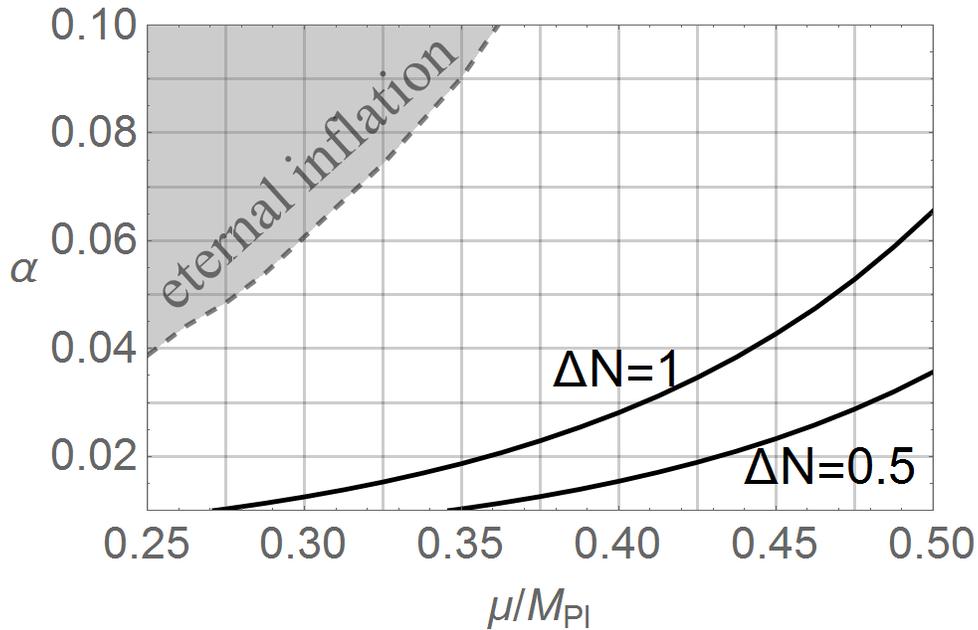}
\end{center}
\caption{
\bl{
A contour plot of $\Delta N$ caused by a hill on top of the 
$R^2$-inflation\bl{-type} potential. The shaded region corresponds to parameters leading to eternal inflation at the patches B.
}
}
\label{Rsquare}
\end{figure}
}
\section{\red{Summary and discussion}}
We have proposed a new mechanism in which primordial perturbations large
enough to produce PBHs are generated while keeping most region\red{s} of the universe
sufficiently homogeneous \cyan{\bl{so} that constraints from CMB distortions can be evaded. In particular, 
our model can explain SMBHs observed at high redshifts by PBHs.}
The basic idea is that each patch of the comoving size corresponding to  
the \red{comoving} Hubble horizon at the time of the PBH formation, after being causally disconnected,
followed one of two different inflationary histories \bl{causing a}
different amount of expansion.
A history followed by a tiny number of patches has more expansion
than the other history followed by most patches.
If this \red{difference \bl{in}} expansion, in terms of the number of $e$-folds, exceeds unity,
\cyan{the minor patches, having experienced more expansion than the major ones, collapse to form PBHs} when they reenter the Hubble horizon.
Since perturbations are \bl{sufficiently} tiny \bl{elsewhere}, nothing special happens
\red{that} might lead to phenomena contradicting with observations.
In particular, no significant CMB distortion is generated in our mechanism
and \red{the} upper bound set by COBE/FIRAS measurement\red{s} can be satisfied.

In our scenario, PBHs of mass $10^{10}M_\odot$\red{, or less considering the growth of these PBHs,} are produced at redshifts $z\: \cyan{\gtrsim}\: 2\times 10^7$.
Thus, this scenario predicts the existence of SMBHs at any redshift
relevant to astrophysical observations, 
\bl{in contrast to potential astrophysical scenarios}
in which
\red{the} number of SMBHs rapidly decreases as the redshift is increased.
If future observations discover SMBHs at \bl{even} higher redshifts,
then our scenario will be \red{a} strong candidate.
On the other hand, if SMBHs turn out to be absent at higher redshifts,
then our scenario \red{will be} disfavored.

\bl{Finally, a few comments are in order. For} the $\chi$ field to take a positive value at around 
$\phi_{\rm{BH}}$ in some patch of $k_{\rm{BH}}$, 
larger regions encompassing that patch must have experienced more "kicks" to the positive direction (see Fig.\ref{fig:bump.eps}). 
This indicates that the spatial distribution of PBHs as the seeds of the SMBHs at high redshifts tend to be clustered in our models, 
\bl{and this clustering} may turn out to be 
inconsistent with observations. One may circumvent this problem by modifying the potential in such a way that the field trajectory is restricted to 
some constant $\tilde{\chi}$ ($\bar{\chi}<\tilde{\chi}<0$) for \bl{$\phi>\phi_c(>\phi_{\mathrm{BH}})$, 
with $\phi_c$ chosen so that spatial clustering can be avoided} and $\tilde{\chi}$ adjusted 
to give an appropriate value of $\beta$, as has been done around (\ref{chibar}). 
This work should be regarded as an existence proof of phenomenological models that can predict PBHs whose mass is sufficiently large to explain SMBHs of $\sim 10^{10}M_\odot$ at high redshifts, and to this end we have introduced two toy models. The potentials we used may appear somewhat contrived, and it would be desirable 
to find simpler and more physically motivated models, that \bl{lead to the same predictions} discussed here.
\section*{ACKNOWLEDGMENTS}
\bl{This work was partially
supported by Grant-in-Aid for JSPS Fellow No. 25.8199 (T.N.),
JSPS Postdoctoral Fellowships for Research Abroad (T.N.), 
JSPS Grant-in-Aid for Young Scientists (B) No. 15K17632 (T.S.),
MEXT Grant-in-Aid for Scientific Research on Innovative Areas 
``New Developments in Astrophysics Through Multi-Messenger 
Observations of Gravitational Wave Sources'' No. 15H00777 (T.S.) and 
``Cosmic Acceleration'' No. 15H05888 (T.S.), 
\sora{the JSPS
Grants-in-Aid for Scientific Research (KAKENHI) 15H02082 (J.Y.).}}
\section*{Appendix \sora{A}: Dependence on primordial non-Gaussianity of $\mu-$distortion constraints on PBHs}
\label{non-Gaussian}
As is discussed in \soutc{section \ref{sec-mu}} \cat{Introduction},
in \cite{Kohri:2014lza} PBHs as the seeds of SMBHs are shown to be constrained by CMB $\mu$ distortions. 
That is, if PBHs with $M_{\rm{PBH}}\gtrsim 10^4M_\odot-10^5M_\odot$ formed by collapse of radiation perturbations provide the seeds of SMBHs, 
CMB spectral distortions larger than observational upper bounds obtained by COBE/FIRAS inevitably arise.
\sky{
Likewise, the formation of PBHs with $M_{\rm{PBH}}\lesssim 10^5M_\odot$ as the potential seeds of SMBHs 
simultaneously leads to an abundant production of dark matter mini-halos (ultracompact mini-halos (UCMHs)) at high redshifts (say, $z\sim 1000$), 
which may emit standard model particles such as photons too 
intensely to be consistent with observed flux obtained by experiments like Fermi (see \cite{Kohri:2014lza}).}
However, \umi{in drawing this conclusion,} primordial perturbations are assumed to be Gaussian, 
and one would expect constraints obtained in \cite{Kohri:2014lza} change for non-Gaussian cases. 
If non-Gaussianity is such that high-$\sigma$ peaks are suppressed, 
then constraints on PBHs from CMB $\mu$ distortions (and potentially from UCMHs, \souta{see the footnote**} \sky{mentioned above}) are even tighter, 
since in this case the dispersion of primordial perturbations for a fixed abundance of PBHs is larger 
than that in a Gaussian case. 
Conversely, if non-Gaussianity is such that high-$\sigma$ peaks are enhanced, 
then $\mu-$distortion constraints on PBHs would be relaxed, and if non-Gaussianity is sufficiently large, 
$\mu-$distortion constraints on PBHs would be completely evaded. This was the essence 
of avoiding CMB distortion constraints to explain most massive SMBHs at high redshifts by PBHs, discussed in this paper. 

\begin{figure}[tb]
  \begin{center}
    \includegraphics[width=100mm]{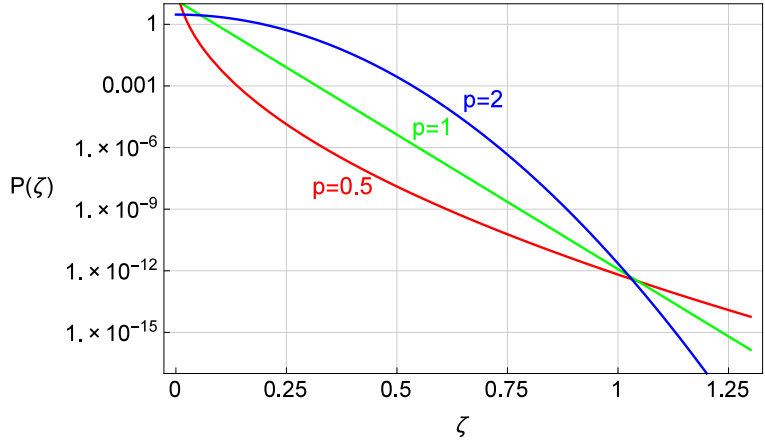}
  \end{center}
  \caption{\red{The PDF of the curvature perturbation $\zeta$ for the same $\beta$ with 
  different values of $p$ of eq. (\ref{pdf2}).}}
  \label{ng}
\end{figure}

In this \cat{a}ppendix we show primordial perturbations have to be \textit{tremendously} non-Gaussian, with high-$\sigma$ peaks enhanced considerably in comparison to a Gaussian case, 
to completely evade constraints on PBHs from CMB distortions, 
adopting the following class of PDFs:
\begin{equation}
P(\zeta)=\frac{1}{2\sqrt{2}\tilde{\sigma} \Gamma\left(1+1/p\right)}\exp \left[-\left(\frac{|\zeta |}{\sqrt{2}\tilde{\sigma}}\right)^p\right],\label{pdf2}
\end{equation}
where $\tilde{\sigma}$ and $p$ are positive. \soutb{That is, here we consider a monotonically decreasing PDF unlike a bimodal one considered in this paper.} 
This \fish{function} satisfies $\int_{-\infty}^\infty P(\zeta)d\zeta =1$ and reduces to a Gaussian PDF when $p=2$. 
If $p<2$ high-$\sigma$ peaks are enhanced compared to the case of $p=2$ and so we restrict our attention to $p<2$ here. 
For general $p$, derivatives at $\zeta=0$ are discontinuous and so 
this PDF is unphysical; however, the purpose of this \fish{a}ppendix is to show that 
$\zeta$ has to be tremendously non-Gaussian for PBHs as the seed of SMBHs to avoid constraints 
from CMB $\mu$ distortion and UCMHs, and this toy model is convenient for that purpose. 
The dispersion is 
\begin{equation}
\sigma^2\equiv \int_{-\infty}^\infty \zeta^2 P(\zeta)d\zeta=\frac{2\Gamma(1+3/p)}{3\Gamma(1+1/p)}\tilde{\sigma}^2,
\end{equation}
where $\Gamma(a)$ is a gamma function. 
In particular, $\sigma=\tilde{\sigma}$ when $p=2$, as it should be. 
The abundance of PBHs is
\begin{equation}
\beta=\int_{\zeta_c}^\infty P(\zeta)d\zeta =\frac{\Gamma(1/p, 2^{-p/2}(\zeta_c/\tilde{\sigma})^p)}
{2p\Gamma(1+1/p)},
\end{equation}
where $\Gamma(a,z)$ is an incomplete gamma function. 
This can be solved for $\tilde{\sigma}$ as 
\begin{equation}
\tilde{\sigma}=\frac{2^{-1/2}\zeta_c}{Q^{-1}(1/p,2\beta)^{1/p}},
\end{equation}
where $Q^{-1}(a,z)$ is the inverse of the regularized incomplete gamma function $Q(a,z)\equiv \Gamma(a,z)/\Gamma(a)$, 
namely, $z=Q^{-1}(a,s)$ if $s=Q(a,z)$. 
The PDF for different values of $p$ for the same $\beta=4\times 10^{-14}$ (see eq.(\ref{beta})) and with $\zeta_c=1$ is shown in Fig. \ref{ng}. 
\begin{figure}[tb]
  \begin{center}
    \includegraphics[width=100mm]{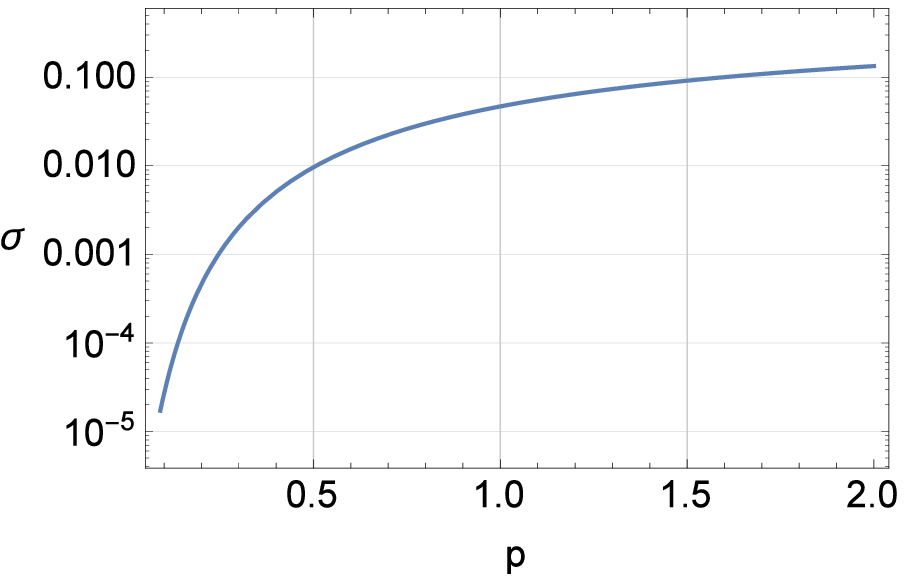}
  \end{center}
  \caption{\red{The root mean square $\sigma$ of $\zeta$ for each $p$,  required to produce a desirable amount of PBHs 
  to explain SMBHs at high redshifts. }}
  \label{psigma}
\end{figure}
Note that all the curves in this figure cross at $\zeta\sim 1$, which is expected since the integral above $\zeta_c\sim 1$ is 
fixed and the dominant contribution to the integral comes from $\zeta\sim 1$. 
In addition, the plot of $\sigma$ as a function of $p$, with $\beta$ fixed to the above value, 
is shown in Fig. \ref{psigma}. 
If $p$ is smaller, the tail of the PDF or the probability of PBH formation is enhanced for fixed $\sigma$, 
and so the value of $\sigma$, required to explain SMBHs at high redshifts by PBHs, is smaller, 
and if $\sigma$ is sufficiently small constraints from CMB $\mu$ distortion and UCMHs can be avoided. 
Let us consider constraints on PBHs obtained from CMB $\mu$ distortion following \cite{Kohri:2014lza}. 
If we assume the following delta-function\bl{-type} power spectrum\soutc{(, which is conservative, see \soutb{the} footnote \ref{conservative},)} leads to a sufficient probability of PBH formation, 
\begin{equation}
{\cal P}_\zeta=\sigma^2 k\delta(k-k_*),\label{deltafunction}
\end{equation}
the $\mu$ distortion generated from this spike is \cite{Chluba:2012we}
\begin{equation}
\mu\simeq 2.2\sigma^2 \left[
\exp\left(-\frac{\hat{k}_*}{5400}\right)
-\exp\left(-\left[\frac{\hat{k}_*}{31.6}\right]^2\right)
\right],
\end{equation}
where $k_*=\hat{k}_*\rm{Mpc}^{-1}$. 
We adopt $\mu_{\rm{upper}}=9\times 10^{-5}$ as a 2$\sigma$ upper limit obtained by COBE/FIRAS \cite{Fixsen:1996nj}. 
The $\mu-$distortion calculated by the above formula as a function of $\hat{k}_*$ for several values of $p$ is shown in Fig. \ref{ngmu}, 
along with the COBE/FIRAS upper limit. 
\cat{This figure is to be compared with Fig. \ref{mu}, which is the corresponding plot for the Gaussian case. }
\begin{figure}[tb]
  \begin{center}
    \includegraphics[width=100mm]{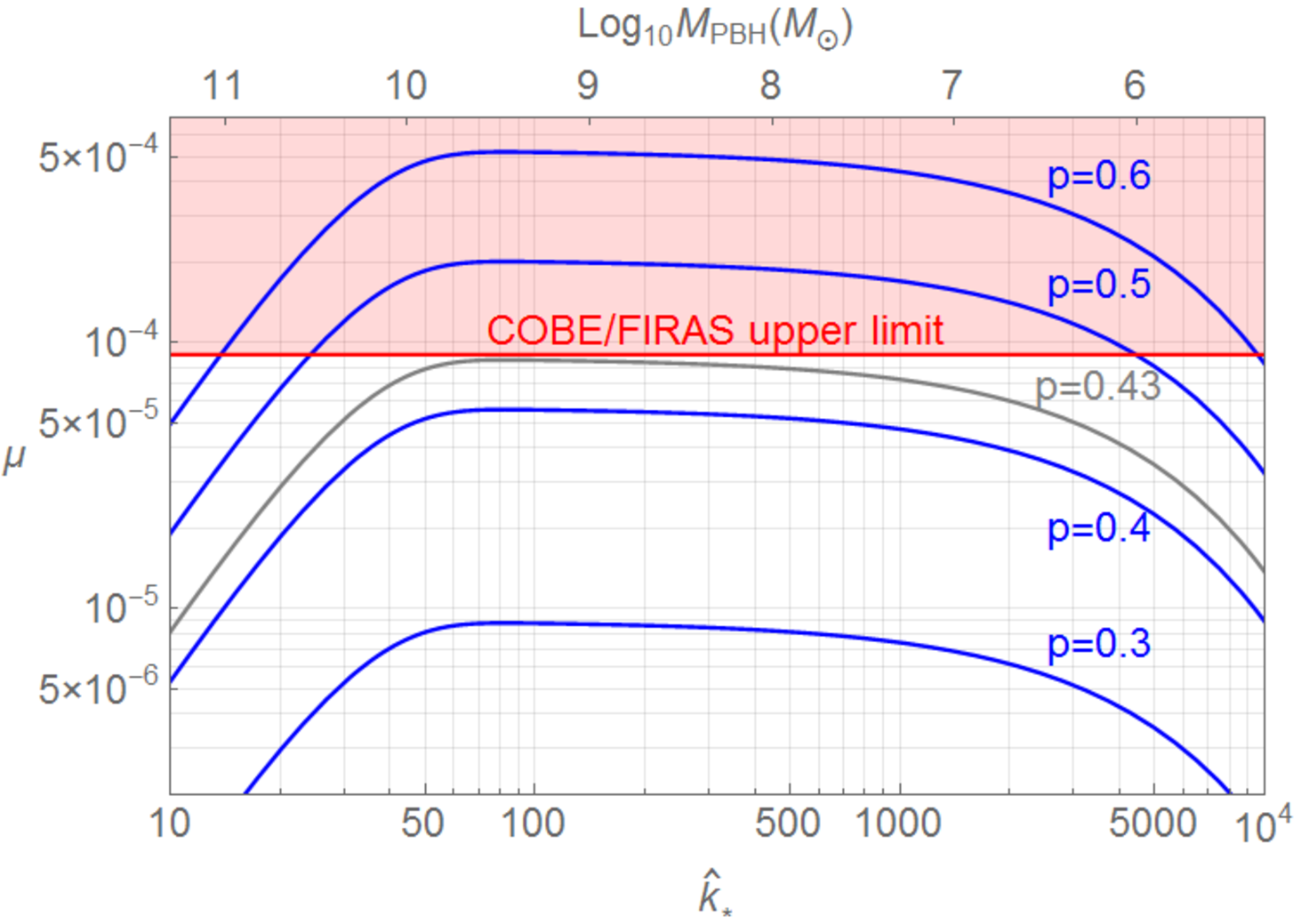}
  \end{center}
  \caption{\bl{The $\mu$ distortion induced by the delta-function-type power spectrum of the curvature perturbation (\ref{deltafunction}) as a function of $\hat{k}_*$, 
  assuming the non-Gaussian PDF (\ref{pdf2}).
  Here $\beta=4\times10^{-14}$ and $\zeta_c=1$ are used. 
  For $0.43<p$, there exists a range of $\hat{k}_*$ that leads to a $\mu$ distortion exceeding the COBE/FIRAS limit.
  }
  }
  \label{ngmu}
\end{figure}
As can be seen from this \cat{Fig. \ref{ngmu}}, if $\mu_{\rm{upper}}\lesssim 2.2\sigma^2$, noting \bl{that} the inside of the square bracket is less than unity, there exists a range of $k_*$ excluded by CMB $\mu$ distortion. 
This condition yields $6.4\times 10^{-3}\lesssim \sigma$ or $0.43\lesssim p$ fixing $\beta$ as above, and if this is satisfied 
approximately a spike in the following range is excluded:
\begin{equation}
31.6\sqrt{-\log\left(1-\frac{\mu_{\rm{upper}}}{2.2\sigma^2}\right)}
\lesssim \hat{k}_*\lesssim -5400 \log\left(\frac{\mu_{\rm{upper}}}{2.2\sigma^2}\right).
\end{equation}
\bl{Using} the following relationship between $k_*$ and the typical mass of PBHs evaluated by the horizon mass when the modes with $k=k_*$ cross the horizon, 
\begin{equation}
M_{\rm{PBH}}=2.2\times 10^{13}\left(\frac{k_*}{1\rm{Mpc}^{-1}}\right)^{-2},
\end{equation}
the above range of $\hat{k}_*$ is translated into the following range of \bl{the mass of PBHs}, excluded by CMB $\mu$ distortion;
\begin{equation}
8\times 10^5M_\odot \left(\log\left(\frac{\mu_{\rm{upper}}}{2.2\sigma^2}\right)\right)^{-2}
\lesssim M_{\rm{PBH}}\lesssim 2\times 10^{10}\left(-\log\left(1-\frac{\mu_{\rm{upper}}}{2.2\sigma^2}\right)\right)^{-1}.\label{range}
\end{equation}
The lower and upper bounds here for each $p$ for the same fixed $\beta$ above are shown in Fig. \ref{excluded}. 
Noting the logarithmic dependence \bl{on $p$} of this mass range, roughly PBHs in $10^6M_\odot\lesssim M_{\rm{PBH}}\lesssim 10^{10}M_{\rm{PBH}}$, 
probably the most important range for PBHs as a candidate for the seeds of SMBHs, are excluded by CMB $\mu$ distortion(, and 
larger PBHs are excluded by CMB $y$ distortions,)
\textit{unless} primordial perturbations are tremendously non-Gaussian ($p\lesssim 0.43$ in the toy model analyzed here), 
with high-$\sigma$ peaks enhanced considerably in comparison to a Gaussian case. 
Smaller PBHs can be potentially constrained by annihilation of dark matter inside UCMHs \cite{Kohri:2014lza}, and 
these potential constraints are also applicable unless primordial perturbations are tremendously non-Gaussian. 
If such a highly non-Gaussian and monotonically decreasing PDF for $0\lesssim \zeta$ can \bl{indeed be} realized in some model of inflation, 
such a model can also explain SMBHs by PBHs, evading constraints from CMB distortions or UCMHs. 
\soutb{Be reminded that in this paper we have constructed models leading to a bimodal PDF (see Fig. \ref{fig:pdf.eps}) to explain SMBHs by PBHs. }
\begin{figure}[tb]
  \begin{center}
    \includegraphics[width=100mm]{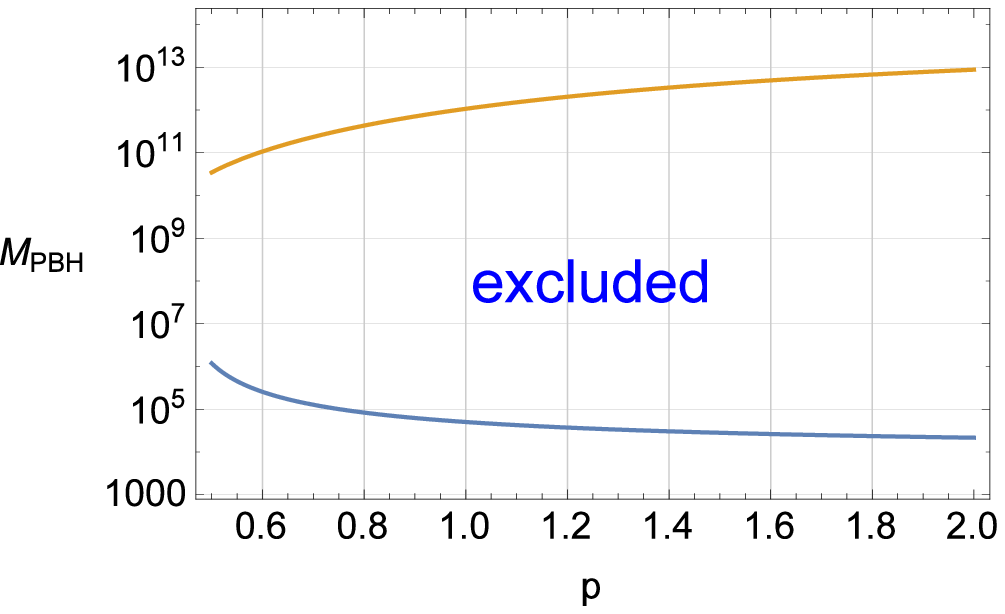}
  \end{center}
  \caption{\red{
  The lower and upper bound of eq. (\ref{range}) for each $p$. 
  The region between the curves corresponds to the mass of PBHs excluded by CMB $\mu$ distortion. 
  }}
  \label{excluded}
\end{figure}
\color{black}
\section*{Appendix B: Mass function of PBHs}
\begin{figure}[t]
\begin{center}
\includegraphics[width=13cm,keepaspectratio,clip]{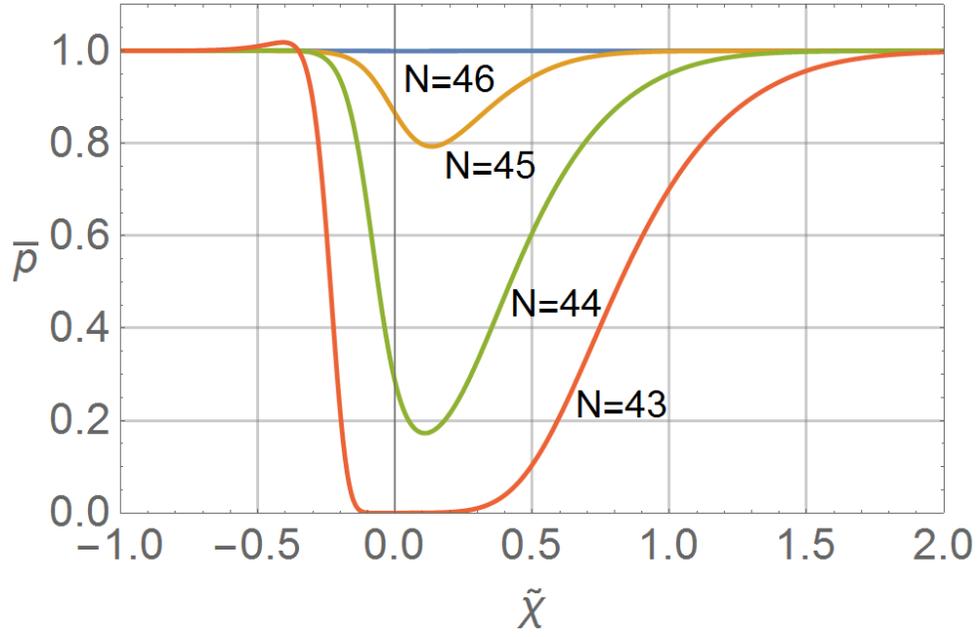}
\end{center}
\caption{
The time evolution of the PDF of the $\chi$ field for the $\phi^2$ potential with $(\alpha,\mu,\tilde{\bar{\chi}},\nu)=(0.06,0.5\cy{M_{\mathrm{Pl}}},-3.6,5.8)$.
}
\label{pdfphisquare}
\end{figure}

\begin{figure}[t]
\begin{center}
\includegraphics[width=13cm,keepaspectratio,clip]{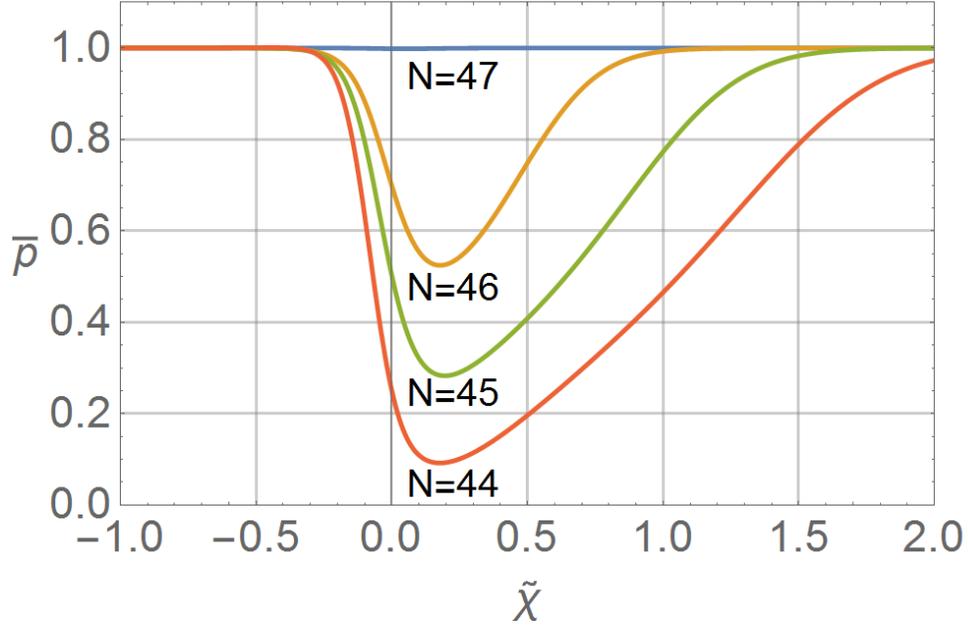}
\end{center}
\caption{
The time evolution of the PDF of the $\chi$ field for the $R^2$-inflation-type potential with $(\alpha,\mu,\tilde{\bar{\chi}},\nu)=(0.02,0.3\cy{M_{\mathrm{Pl}}},-3.6,6.6)$.
}
\label{pdfRsquare}
\end{figure}

Here we calculate the mass function of PBHs in our model by solving the Fokker-Planck equation for the time evolution of the PDF $p(t,\chi)$ of the $\chi$ field. 
To this end we replace the step function $\theta(\chi)$ in eq. (\ref{lag}) by a hyperbolic tangent function as follows:
\begin{equation}
\theta(\chi)\rightarrow T(\chi)\equiv\frac{1}{2}\left[1+\tanh\left(\frac{\chi}{\Delta_\chi}\right)\right],\label{replacement}
\end{equation}
where $\Delta_\chi$ is a positive parameter.
The Fokker-Planck equation for $p(t,\chi)$ is\footnote{Strictly speaking the evolution of $\phi$ is affected by the motion of $\chi$,
but here we investigate the motion of $\chi$ when the effects of the hill on the evolution of $\phi$ are negligible, so the evolutions of $\phi$ and $\chi$ would be  separately treated safely, as is done in this appendix. 
}
\begin{equation}
\frac{\partial p(t,\chi)}{\partial t}=
\frac{V(\phi(t))v(\phi(t))}{3H(t)}\frac{\partial}{\partial \chi}\left[\frac{\partial T(\chi)}{\partial \chi}p(t,\chi)\right]+\frac{H^3(t)}{8\pi^2}\frac{\partial^2p(t,\chi)}{\partial \chi^2}.
\end{equation}
In terms of the $e$-folds $N$, this reads
\begin{equation}
-\frac{\partial p(N,\chi)}{\partial N}=
\frac{V(\phi(N))v(\phi(N))}{3H^2(N)}\frac{\partial}{\partial \chi}\left[\frac{\partial T(\chi)}{\partial \chi}p(N,\chi)\right]+\frac{H^2(N)}{8\pi^2}\frac{\partial^2p(N,\chi)}{\partial \chi^2}.
\end{equation}
Notice that \fish{the} first term of the right-hand side represents the effect of the gap at $\chi\sim 0,$ and
the ratio of the second term to the first term is roughly given by $R$ introduced in (\ref{condition}). That is, the first term becomes important when $R\lesssim 1.$
Let us rewrite the above equation using quantities normalized by $H_{\rm{obs}}=H(t_{\mathrm{obs}})$, denoted with a tilde (e.g. $\tilde{H}=H/H_{\rm{obs}}$). For the case of the $\phi^2$ potential, we obtain
\begin{equation}
-\frac{\partial p(N,\tilde{\chi})}{\partial N}
=f_1(N,\tilde{\chi})p(N,\tilde{\chi})
+f_2(N,\tilde{\chi})\frac{\partial p(N,\tilde{\chi})}{\partial \tilde{\chi}}
+\frac{\tilde{H}^2}{8\pi^2}\frac{\partial^2p(N,\tilde{\chi})}{\partial \tilde{\chi}^2},\label{fp1}
\end{equation}
where
\begin{equation}
f_1(N,\tilde{\chi})\equiv\frac{\tilde{m}^2\tilde{\phi}^2(N)v(\tilde{\phi}(N))}{12\tilde{H}^2(N)\tilde{\Delta_\chi}}
\frac{\partial}{\partial \tilde{\chi}}\left[\sech^2\left(\frac{\tilde{\chi}}{\tilde{\Delta_\chi}}\right)\right],\quad f_2(N,\tilde{\chi})\equiv \frac{\partial f_1(N,\tilde{\chi})}{\partial \tilde{\chi}}.
\end{equation}
For the case of the $R^2$-inflation-type potential, 
\begin{equation}
f_1(N,\tilde{\chi})\equiv\frac{\tilde{M}^2\tilde{M}_{\mathrm{Pl}}^2v(\tilde{\phi}(N))}{8\tilde{H}^2(N)\tilde{\Delta_\chi}}
\left[1-\exp\left(-\sqrt{\frac{2}{3}}\frac{\tilde{\phi}(N)}{\tilde{M}_{\mathrm{Pl}}}\right)\right]^2
\frac{\partial}{\partial \tilde{\chi}}\left[\sech^2\left(\frac{\tilde{\chi}}{\tilde{\Delta_\chi}}\right)\right],
\end{equation}
and $f_2\equiv\partial f_1/\partial \tilde{\chi}$.
We fix $\tilde{\Delta}_{\chi}=1/2\pi$\cy{, 
a typical distance $\chi$ travels over one Hubble time\footnote{\cy{
The smooth transition of the potential, specified by (\ref{replacement}), introduces patches where $10^{-5}<\zeta<1$.\soutb{, unlike the situation where the transition is steep, leading to the bimodal PDF depicted in Fig. \ref{fig:pdf.eps} in the main text.} Though the fraction of such patches is larger than that of patches B forming PBHs, it is significantly smaller than unity with this choice of $\tilde{\Delta}_{\chi}$, so the substantial global $\mu$ distortion can still be avoided. 
}
}.}
Since \cy{$\tilde{H}\sim {\cal O}(1)$ and $\tilde{m}\tilde{\phi},\tilde{M}\tilde{M}_{\mathrm{Pl}}\gg 1$}, the first and second terms above indeed become important when $v\ll \alpha$, far from the location of the center of the hill at $\phi=\phi_0.$
We denote by $p_0$ the solution when the hill or the gap at $\chi\sim 0$ is absent ($\alpha=0$), satisfying a diffusion equation with a (weakly) time-dependent diffusion coefficient, 
\begin{equation}
-\frac{\partial p_0(N,\tilde{\chi})}{\partial N}
=\frac{\tilde{H}^2(N)}{8\pi^2}\frac{\partial^2p_0(N,\tilde{\chi})}{\partial\tilde{\chi}^2},\label{fp2}
\end{equation}
\soutb{the solution of which}\fish{whose solution} is
\begin{equation}
p_0(N,\tilde{\chi})=\frac{1}{\sqrt{2\pi\sigma_{\tilde{\chi}}^2(N)}}\exp\left[-\frac{(\tilde{\chi}-\tilde{\bar{\chi}})^2}{2\sigma_{\tilde{\chi}}^2(N)}\right], 
\quad \sigma_{\tilde{\chi}}^2(N)\equiv\frac{1}{(2\pi)^2}\int_N^{N_{\mathrm{obs}}}\tilde{H}^2(N)dN.
\end{equation}
Let us introduce $\bar{p}(N,\tilde{\chi})\equiv p(N,\tilde{\chi})/p_0(N,\tilde{\chi})$, then from 
(\ref{fp1}) and (\ref{fp2}) its time evolution is determined by 
\begin{equation}
-\frac{\partial \bar{p}(N,\tilde{\chi})}{\partial N}=f_3(N,\tilde{\chi})\bar{p}(N,\tilde{\chi})
+f_4(N,\tilde{\chi})\frac{\partial \bar{p}(N,\tilde{\chi})}{\partial\tilde{\chi}}+f_5(N,\tilde{\chi})\frac{\partial^2\bar{p}(N,\tilde{\chi})}{\partial\tilde{\chi}^2},
\end{equation}
where
\begin{equation}
f_3\equiv f_1+f_2\frac{\partial(\log p_0)}{\partial\tilde{\chi}},\quad f_4\equiv f_2+\frac{\tilde{H}^2}{4\pi^2}\frac{\partial(\log p_0)}{\partial\tilde{\chi}},\quad f_5\equiv \frac{\tilde{H}^2}{8\pi^2}.
\end{equation}
 We solve the above differential equation with the initial condition $\bar{p}(N_\mathrm{i},\tilde{\chi})=1$, with $N_{\mathrm{i}}$ lying between $N_{\mathrm{BH}}$ and $N_{\mathrm{obs}}$. It is taken to be sufficiently large so that the effect of gap is still negligible at $N_{\mathrm{i}}$. 
The boundary conditions are $\bar{p}(N,\pm \infty)=1$\footnote{\cy{
The boundary condition $\bar{p}(N,\infty)=1$ may seem less obvious than 
$\bar{p}(N,-\infty)=1,$ but the probability at $\chi=\infty$ is mostly determined by diffusion from the initial position $\bar{\chi}$ before the gap becomes important, so it would be sufficiently accurate as long as the gap is negligible close to $N_{\mathrm{obs}}.$}
}. 
Evidently, when $\alpha=0$ ($f_1=f_2=0$) the solution is $\bar{p}=1$, as it should.
The simplest finite difference method would suffice, namely, 
\begin{equation}
-\frac{\bar{p}_i^{n+1}-\bar{p}_i^n}{dN}
=f_{3,i}^n\bar{p}_i^n+f_{4,i}^n\frac{\bar{p}_{i+1}^n-\bar{p}_{i-1}^n}{2d\tilde{\chi}}+f_{5,i}^n\frac{\bar{p}_{i+1}^n-2\bar{p}_i^n+\bar{p}^n_{i-1}}{d\tilde{\chi}^2}.
\end{equation}
We take $d\tilde{\chi}=0.0075$. For fixed $d\tilde{\chi}$, $|dN|$ has to be sufficiently small to avoid numerical instability (Courant-Friedrichs-Lewy Condition), and we take $dN=-0.002.$ As illustrations, the time evolution of $\bar{p}$ is shown in Figs. \ref{pdfphisquare} and \ref{pdfRsquare} for the $\phi^2$ potential 
with $(\alpha,\mu,\tilde{\bar{\chi}},\nu)=(0.06,0.5\cy{M_{\mathrm{Pl}}},-3.6,5.8)$ and the $R^2$-inflation-type potential with $(\alpha,\mu,\tilde{\bar{\chi}},\nu)=(0.02,0.3\cy{M_{\mathrm{Pl}}},-3.6,6.6)$. These values of $\alpha$ and $\mu$ can realize $\Delta N$ larger than unity from Figs \ref{phisquare} and \ref{Rsquare}, necessary for PBH formation at patches B. In addition, the above values of $\tilde{\bar{\chi}}$ and $\nu$ are chosen so that the right amount of PBHs of the desired mass is realized, discussed shortly. 
The probability is depleted around the slope at $\chi\sim 0,$ with $\chi$ pushed back toward the negative-$\chi$ region, where $\bar{p}$ becomes slightly larger than unity. This increase in the probability in the left of the slope is only barely noticeable in Fig. \ref{pdfphisquare}, since the probability there is mostly determined by the influx of larger probability from the left, and the effect of the slope there is basically negligible. In contrast, 
the effect of the depleted probability at around the gap gradually propagates toward the positive-$\chi$ region more noticeably, 
since from the point of the right of the slope the crucial supply of probability from the left is cut off as the height of the gap increases.


\begin{figure}[t]
\begin{center}
\includegraphics[width=13cm,keepaspectratio,clip]{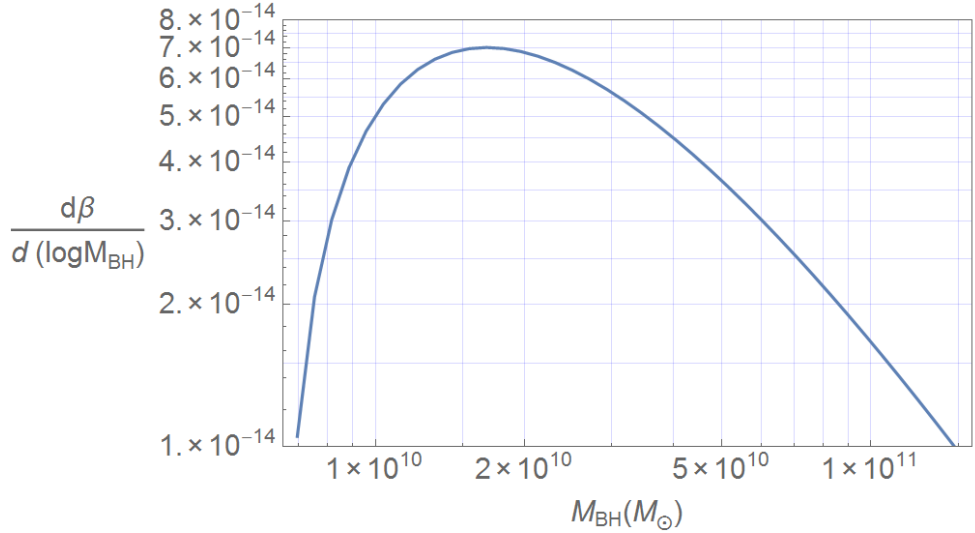}
\end{center}
\caption{
The mass function of PBHs for the $\phi^2$ potential with $(\alpha,\mu,\tilde{\bar{\chi}},\nu)=(0.06,0.5\cy{M_{\mathrm{Pl}}},-3.6,5.8)$.
}
\label{massfunctionphisquare}
\end{figure}

\begin{figure}[t]
\begin{center}
\includegraphics[width=13cm,keepaspectratio,clip]{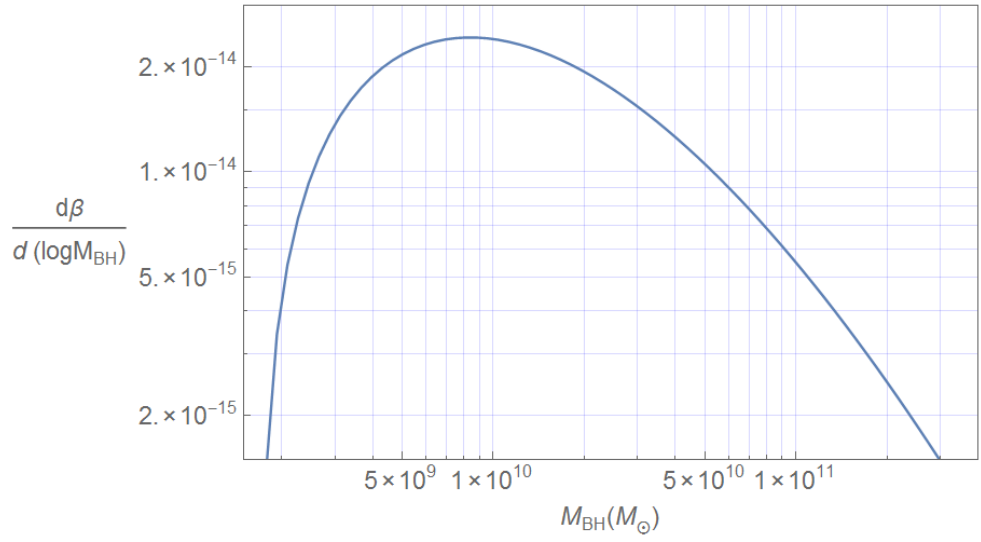}
\end{center}
\caption{
The mass function of PBHs for the $R^2$-inflation-type potential with $(\alpha,\mu,\tilde{\bar{\chi}},\nu)=(0.02,0.3\cy{M_{\mathrm{Pl}}},-3.6,6.6)$.
}
\label{massfunctionRsquare}
\end{figure}

\sky{The \fish{PDF} of $\chi$ is related to the mass function of PBHs as follows. First let us introduce\footnote{
The choice of the lower bound of the integration here \souta{won't} \sky{will not} affect the conclusion of this appendix.
\souta{See also the footnote**.}}
\begin{equation}
\beta(N)\equiv \int_{H_{\mathrm{obs}}/2\pi}^\infty p(N,\chi)d\chi,
\end{equation}
which is the fraction of patches in which $\chi>H_{\mathrm{obs}}/2\pi$ at an $e$-fold $N$. 
Then $(d\beta/d\log N)d\log N$ 
is approximately the fraction of patches in which $\chi$ crosses $\chi=H_{\mathrm{obs}}/2\pi$ from left to right during the interval $(N,eN)$. Strictly speaking that fraction is slightly larger than $(d\beta/d\log N)d\log N$ due to the nonzero fraction of patches in which $\chi$ crosses $\chi=H_{\mathrm{obs}}/2\pi$ from \textit{right to left} during the same interval, but such fraction is negligible unless the height of the gap is sizable. The patches in which $\chi$ crosses $H_{\mathrm{obs}}/2\pi$ from left to right during $(N,eN)$  collapse to PBHs (basically, see a discussion after (\ref{spectrum})) whose mass is related to that $e$-fold by (\ref{kbh}) and (\ref{log}). 
Then the volume fraction of PBHs whose mass lies between $(M_{\mathrm{BH}},eM_{\mathrm{BH}})$ is approximately 
\begin{equation}
\frac{d\beta(N(M_{\mathrm{BH}}))}{d(\log M_{\mathrm{BH}})}d(\log M_{\mathrm{BH}})
=\frac{1}{2}\frac{d\beta(N)}{dN}\bigg|_{N=N(M_{\mathrm{BH}})}d(\log M_{\mathrm{BH}})
\end{equation}
where (\ref{kbh}) and (\ref{log}) have been used to obtain the equality. }The mass function for the $\phi^2$ potential and the $R^2-$inflation-type potential is shown in Figs. \ref{massfunctionphisquare} and \ref{massfunctionRsquare}. 
These show that it is indeed possible to choose the model parameters to realize the PBH mass function with the right abundance (see (\ref{beta})) and at the right mass, here taken to be $M_{\mathrm{BH}}\sim 10^{10}M_\odot.$ The width $\Delta M_{\mathrm{BH}}$ of the mass function turns out to be $\Delta M_{\mathrm{BH}}/M_{\mathrm{BH}}\sim {\cal O}(1).$  
\color{black}
\clearpage
\bibliographystyle{h-physrev}
\bibliography{bib.bib}
\end{document}